\documentclass[epjST]{svjour}

\usepackage{amsmath}
\usepackage{graphics}
\usepackage{color}
\usepackage[sort&compress,square,comma,numbers]{natbib}
\usepackage{subfig}
\usepackage{upgreek}
\usepackage{tikz}
\usepackage{pgfplots}
 \pgfplotsset{compat=1.3}
 \usetikzlibrary{external}
\newcommand{\tension}{\zeta}
\newcommand{\sound}{c_{\text s}}

\begin{document}

\title{Numerical simulations of complex fluid-fluid interface dynamics}
\author{T.~Kr\"uger\inst{3,1}\fnmsep\thanks{\email{t.krueger@ucl.ac.uk}} \and S.~Frijters\inst{1}\fnmsep\thanks{\email{s.c.j.frijters@tue.nl}} \and F.~G\"unther\inst{1}\fnmsep\thanks{\email{f.s.guenther@tue.nl}} \and B.~Kaoui\inst{1}\fnmsep\thanks{\email{b.kaoui@tue.nl}} \and J.~Harting\inst{1,2}\fnmsep\thanks{\email{j.harting@tue.nl}}}
\institute{Department of Applied Physics, Eindhoven University of Technology, P.O. Box 513, NL-5600MB Eindhoven, The Netherlands \and Institute for Computational Physics, University of Stuttgart, Pfaffenwaldring 27, D-70569 Stuttgart, Germany \and Centre for Computational Science, University College London, 20 Gordon Street, London WC1H 0AJ, United Kingdom}
\abstract{Interfaces between two fluids are ubiquitous and of special
importance for industrial applications, e.g., stabilisation of
emulsions. The dynamics of fluid-fluid interfaces is difficult to study because
these interfaces are usually deformable and their shapes are not known a
priori. Since experiments do not provide access to all observables of
interest, computer simulations pose attractive alternatives to gain
insight into the physics of interfaces. In the present article, we
restrict ourselves to systems with dimensions comparable to the lateral
interface extensions. We provide a critical discussion of three numerical
schemes coupled to the lattice Boltzmann method as a solver for the hydrodynamics of the problem:
(a) the immersed boundary method for the simulation of vesicles and
capsules, the Shan-Chen pseudopotential approach for multi-component
fluids in combination with (b) an additional advection-diffusion component
for surfactant modelling and (c) a molecular dynamics algorithm for the
simulation of nanoparticles acting as emulsifiers.}

\maketitle

\section{Introduction}
\label{sec:intro}

Interfaces are ubiquitous in soft matter systems and appear in various
shapes and sizes. Prominent examples are vesicles (closed membrane defined by a bilayer of
phospholipid molecules \cite{seifert_configurations_1997}), capsules
(closed polymeric membrane \cite{pozrikidis2003modeling}), and all kinds of biological
cells \cite{pozrikidis2003modeling}. In these cases, the interface is an
additional material whose constitutive behavior has to be specified. For
example, vesicle membranes are incompressible and viscous
\cite{seifert_configurations_1997} whereas capsule membranes resist shear
and can be compressible \cite{gompper2008soft}. Understanding the dynamics
of membranes is important for disease detection by measuring mechanical
properties of living cell membranes \cite{suresh2005connections}, targeted
drug delivery \cite{battaglia2012lipid}, and predicting the viscosity of
biofluids, such as blood \cite{fedosov2011predicting}. Typical
applications are lab-on-chip devices for particle identification and
separation \cite{hou2010deformability}.

Other classes of fluid-fluid interfaces can be found in emulsions (liquid drops
suspended in another liquid), foams (gas bubbles separated by thin liquid
films), and liquid aerosols (liquid drops in gas) where at least two immiscible
fluid phases are mixed. The interface is then defined by the common boundaries
of the phases. Emulsions are of central importance for food processing (e.g.,
milk, salad dressings) \cite{silva2011nanoemulsions}, cosmetics and
pharmaceutics (e.g., lotions, vaccines, disinfection) \cite{puglia2012lipid},
and enhanced oil recovery \cite{hirasaki2011recent}.


Emulsions and foams are usually unstable. Drops and bubbles tend to
coalesce gradually, which reduces the interfacial free energy. Conversely,
droplet breakup can be triggered by shearing the system and pumping energy
into it. This is of practical importance for mechanical emulsification
\cite{jafari2008re}. A significant amount of research has been focused on
the question how to stabilize emulsions, thus avoiding or at least
retarding coalescence. The traditional approach is to add surfactants to
an emulsion \cite{bib:jens-harvey-chin-venturoli-coveney:2005}. Surfactants are amphiphilic molecules for which it is
energetically favorable to accumulate at the interface, which in turn
leads to a decrease of surface tension and prevents demixing. Under some
circumstances, it is possible to find so-called mesophases where the
phases percolate, penetrate each other, and are separated by a interfacial
surfactant monolayer. These mesophases can be periodic such as the diamond
or gyroid minimal surfaces
\cite{bib:jens-harvey-chin-venturoli-coveney:2005,bib:jens-harvey-chin-coveney:2004,bib:gompper-schick:1994}.

An alternative route to emulsion stabilization is by using solid
nanoparticles: these particles also accumulate at the interface where they
replace segments of fluid-fluid interface by fluid-particle interfaces and
thus lower the interfacial free energy. However, the physical mechanism is
different from that of stabilization by surfactants, and the surface
tension is unchanged by the nanoparticles \cite{binks2002particles,
tcholakova2008comparison, frijters2012effects}. For
nanoparticle-stabilized emulsions, one distinguishes between the so-called
``Pickering emulsions'' \cite{bib:ramsden:1903,pickering1907cxcvi} and ``bijels''
(bicontinuous interfacially jammed emulsion gels)
\cite{stratford2005colloidal, herzig2007bicontinuous}. The former is an
emulsion of discrete droplets in a continuous liquid. In the latter, both
phases are continuously distributed. Besides their primary task to act as
emulsifiers, the nanoparticles may additionally be designed as, for
example, ferromagnetic \cite{kim2010bijels} or Janus particles
\cite{binks2001particles}.

The macroscopic dynamics of complex fluids strongly depends on the microscopic
properties of interfaces present in these systems. The reason is that the ratio
between interface surface and bulk volume is usually large. The interfaces act
as additional boundary conditions restricting the dynamics of the fluids.
Quantities of interest are the interface's shear and dilatational viscosities
and the surface dilatational modulus if the interface is compressible (which is
--- in contrast to bulk rheology --- often the case). Interface models usually
have to obey mass, momentum, and energy conservation. If isothermal or athermal
situations are sufficient, the energy conservation can be relaxed. For a review
of experimental methods characterizing interfaces in soft matter and finding
thermodynamically consistent constitutive laws, we refer to
\cite{sagis2011dynamic}.

Fluid-fluid interface problems are hard to solve analytically since the
interface is usually deformable and its shape not known a priori. Therefore,
the interface dynamics is fully coupled to that of the ambient phases.
Interface deformations are rarely small and linear, which results in complex
and time-dependent boundary conditions. Analytical solutions are only available
for academic cases (e.g., \cite{barths-biesel_time-dependent_1981,
wetzel2001droplet}). These circumstances call for numerical methods and
computer simulations which can also provide access to observables not traceable
in experiments such as local interface curvature or fluid and interface
stresses. Therefore, computer simulations can be used to complement
experiments.

Engineering applications call for a better understanding of the link between
the microscale and the emergent macroscopic behavior of these systems. For
example, it is an open question how to generally predict the macroscopic
behavior (such as the effective viscosity) of a complex fluid based on its
microscopic properties (e.g., surface tension or dilation modulus of the
interface, wettability of the nanoparticles) and how to manipulate these
properties in order to tailor new materials with specific macroscopic behavior
\cite{larson1999structure, benzi2010herschel, pal2011rheology}. What are the
underlying physical mechanisms for emulsion stabilization due to surfactants
and nanoparticles, and how can their costs and environmental sustainability be
optimized? How can emulsions with a desired droplet size distribution be
produced? How do drops, vesicles, and capsules behave in non-trivial flow
geometries, and how can they be separated based on their interfacial
properties?

In this article, we focus on vesicles, capsules, and emulsions in systems with
dimensions comparable to the lateral extension of the interfaces, e.g., a
volume containing ${\cal O}(1-100)$ objects. At this scale, fluid dynamics is
dominated by effects due to viscosity, surface tension or elasticity, and
inertia. Gravity is usually not relevant in microfluidics because the capillary
length for water droplets in air is typically of the order of a few
millimeters. Due to the small system size, the total amount of energy
dissipated is small and heat is almost instantaneously transported out of the
system by heat conduction.
It is therefore an excellent approximation to assume isothermal systems with
constant viscosity and surface tension. The relative strength of the three
major contributions (viscosity, surface tension, inertia) can be described by
dimensionless parameters, such as the capillary number (Ca, viscous stress vs.
surface tension or elasticity), the Reynolds number (Re, inertial vs. viscous
stresses), or the Weber number (We, inertial stress vs. surface tension). Note
that the three numbers obey $\text{We} \propto \text{Re} \times \text{Ca}$ and
are not independent.

\begin{figure}
 \centering
 \subfloat[continuum]{
  \includegraphics{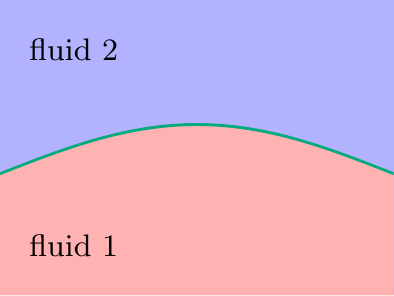}
 }
 \subfloat[mesoscopic]{
  \includegraphics{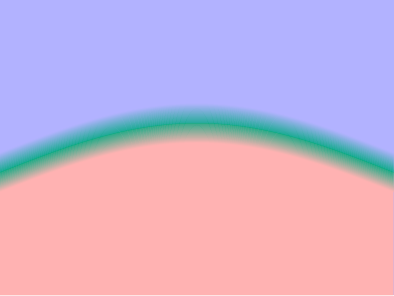}
 }
 \subfloat[discrete]{
  \includegraphics{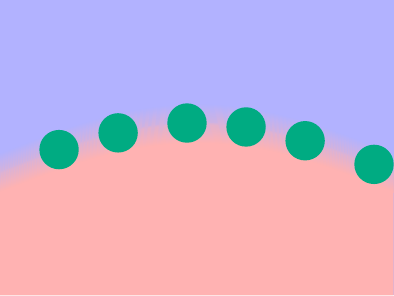}
 }
 \caption{\label{fig:coarse_graining} Sketch of coarse-graining levels for an interface between two fluids (fluids 1 and 2). In panel (a), the interface (green line) is explicitly tracked in a Lagrangian manner, and the interface properties have to be provided in form of a constitutive law. There is no direct interaction between the fluid species. The substructure of the interface is not known (continuum picture). For a surfactant-stabilised droplet as shown in panel (b), the interface is diffuse with an additional dynamical surfactant field (green). The local mutual interactions between the surfactant and the two fluid components have to be specified, but the microscopic surfactant properties are not required (mesoscopic picture). In panel (c), the nanoparticles (green) are explicitly resolved (discrete picture). Their individual motions are tracked, and their geometrical and interaction properties have to be specified (e.g., shape, contact angles). The interaction of the fluid species has to be defined as well. For (b) and (c), the choice of the mutual interactions leads to an emergent macroscopic interface behaviour.}
\end{figure}

In the current article, the fluid phases are treated as Newtonian fluids
simulated by the lattice Boltzmann method (detailed in section
\ref{sec:numerical}). Depending on the physical conditions and the desired
level of coarse-graining, it may be sufficient to treat the interface as an
effective two-dimensional material obeying a well-defined constitutive law or
it may be necessary to resolve the substructure of the interface (cf.\ fig.\
\ref{fig:coarse_graining}). We consider three cases: In the continuum
description for vesicle- and capsule- membranes, the immersed boundary method
is coupled to the single-component lattice Boltzmann method (section
\ref{sec:macroscopic}). The lattice Boltzmann Shan-Chen multi-component model
is the basis for the remaining two cases. In the first, a surfactant
concentration field is employed (mesoscopic approach, section
\ref{sec:mesoscopic}), in the second, explicitly resolved solid nanoparticles
interact with the fluid phases (discrete approach, section
\ref{sec:microscopic}). In section \ref{sec:results}, some exemplary results
from simulations of a vesicle under shear flow, separating red blood cells and
platelets, surfactant stabilized interfaces, and particle stabilized interfaces
are presented. Finally, we conclude in section \ref{sec:conclusions}.

\section{Numerical approaches}
\label{sec:numerical}
The physical problem consists in capturing the dynamics of an interface
separating distinct fluids. Most of the macroscopic properties of soft
matter systems, for example the rheology of an emulsion, result from the
reorganisation and structuring of interfaces at the microscale. In the
simplest situation, there are two different fluids separated by an
interface. Mathematically, this corresponds to two domains separated by a
free-moving boundary. There are two classes of numerical approaches to
track the motion of an interface: (i) either \textit{front tracking}
(e.g., the boundary integral method or the immersed boundary method) in which the
motion of marker points attached to the interface is tracked, or (ii)
\textit{front capturing} (e.g., the phase field method or level set
methods) where a scalar field is used as an order parameter, indicating
the composition of the fluid at a given point. For example, the order
parameter may take the values $0$ and $1$ in the bulk regions of fluid
component A and B, respectively, and it is smoothly varying in the
intermediate regions. The interface is located where the scalar field
assumes a well-defined value, e.g., $0.5$ in the above example. The time
evolution of the order parameter is governed by an advection-diffusion
equation coupled to the fluid flow. In both approaches (front tracking and
capturing), the momentum of the incompressible Newtonian fluid obeys the
Navier-Stokes (NS) equations,
\begin{equation}
\rho \left(\frac{\partial {\bf u}}{\partial t} + {\bf u} \cdot \nabla {\bf u}\right) = -\nabla p + \eta \nabla^2 {\bf u},\quad \nabla \cdot {\bf u} = 0.
\end{equation}
The mass density and dynamic viscosity of the fluid are denoted by $\rho$ and
$\eta$; ${\bf u}$ and $p$ are its velocity and pressure fields and $t$ denotes
the time. For the solution of the NS equations, boundary conditions at the
interface have to be considered. However, the location of the interface is on
its own an additional unknown quantity. Its motion results from its
hydrodynamical interaction with the surrounding fluid. Therefore, to obtain the
dynamics of the interface, information about the flow is required. In the
present article, instead of solving the NS equations directly, we alternatively
use a mesoscopic approach: the \textit{lattice Boltzmann method} (LBM).

\begin{figure}
 \subfloat[lattice and IBM interpolation]{
  \includegraphics[height=4.5cm]{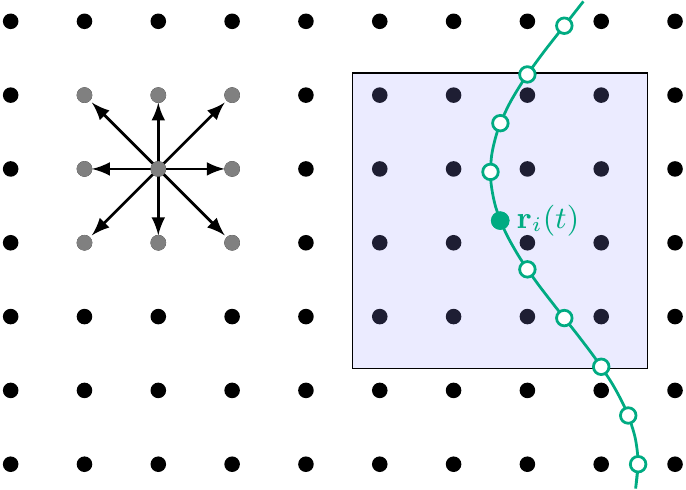}
 } \hfill
 \subfloat[solid moving boundaries]{
  \includegraphics[height=4.5cm]{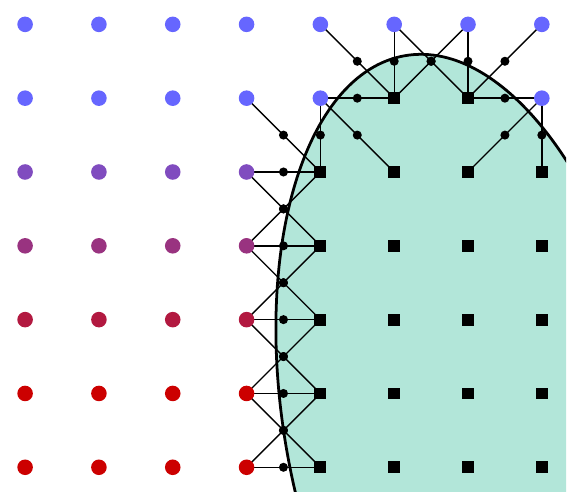}
 }
 \caption{\label{fig:lattice_etc} Two-dimensional illustrations of the lattice Boltzmann propagation, immersed boundary interpolation, and bounce-back. (a) During propagation, the populations $f_i$ move to their next neighbors (gray lattice sites and arrows). Within the immersed boundary method, an interface node (here: node $i$ at position ${\bf r}_i(t)$) is coupled to the single-component lattice fluid. The range of the interpolation stencil (here: $4 \times 4$) is denoted by the square region. All lattice sites within this region have to be considered during velocity interpolation and force spreading. (b) An ellipsoidal nanoparticle is located at a fluid-fluid interface. One fluid component is indicated by red, the other by blue circles. The colour gradient illustrates the interface region. Squares denote lattice sites inside the particle. Populations are bounced back at points (shown as small circles) located half-way between neighbouring fluid and particle sites. This gives rise to an effective staircase description of the particle shape.}
\end{figure}

The LBM has gained popularity among scientists and engineers because of its relatively straightforward implementation compared to other approaches such as the finite element method (for reviews, see \cite{Succi2001,Sukop2006,Aidun2010}). In the LBM, the fluid is considered as a cluster of pseudo-particles that move on a lattice under the action of external forces. To each pseudo-particle is associated a distribution function $f_i$, the main quantity in the LBM. It gives the probability to find a pseudo-particle at a position ${\bf r}$ with a velocity in direction ${\bf e}_i$. In the LBM, both the position and velocity spaces are discretised: $\Delta x$ is the grid spacing, and ${\bf e}_i$ are the discretised velocity directions. There are different types of lattices available: For 2D simulations, we use the so-called D2Q9 lattice with nine velocity directions (see also Fig.~\ref{fig:lattice_etc}(a)); for 3D simulations, the D3Q19 lattice with nineteen velocity directions is employed \cite{qian_lattice_1992}. The time evolution of $f_i$ is governed by the so-called lattice Boltzmann equation,
\begin{equation}
\label{eq:lbe}
f_i({\bf r} + {\bf e}_i \Delta t, t + \Delta t) - f_i({\bf r}, t) = \Omega_i,
\end{equation}
where $\Delta t$ is the discrete time step. The lattice Boltzmann equation
(\ref{eq:lbe}) consists of two parts: (i) the advection part (left-hand side)
and (ii) the collision part (right-hand side) with collision operator
$\Omega_i$ specifying the collision rate between the fluid pseudo-particles. It
can be approximated by the Bhatnagar-Gross-Krook (BGK) operator, $\Omega_i =
-(f_i - f_i^{\text{eq}})/ \tau$, which describes the relaxation of $f_i$
towards its local equilibrium, $f_i^{\text{eq}}$, on a time scale $\tau$. The
relaxation time is related to the macroscopic dynamical viscosity $\eta$ via
$\eta = \rho \sound^2\frac{\Delta x^2}{\Delta t}\left(\tau - \frac{1}{2}
\right)$, where $\sound = 1 / \sqrt{3}$ is the lattice speed of sound. The
equilibrium distribution is given by a truncated Maxwell-Boltzmann
distribution,
\begin{equation}
\label{eq:equilibrium}
f_i^{\text{eq}} (\rho, {\bf u}) = \omega_i \rho \left[1 + \frac{{\bf c}_i \cdot {\bf u}}{\sound^2} + \frac{({\bf c}_i \cdot {\bf u})^2}{2\sound^4} - \frac{{\bf u} \cdot {\bf u}}{\sound^2}\right],
\end{equation}
where the $\omega _i$ are weight factors resulting from the velocity space
discretisation. The hydrodynamical macroscopic quantities are computed
using the first and the second moments of $f_i$: (i) the local pressure $p
= \rho \sound^2 = \sound^2 \sum_i f_i$ and (ii) the local velocity ${\bf
u} = \sum_i f_i {\bf e}_i / \rho$. Even though a conversion of the units
to SI units is straightforward, it is common practise to report all quantities in lattice units. Non-dimensionalisation is possible by choosing suitable combinations of $\Delta x$, $\Delta t$, and the fluid density.

The LBM can also handle multi-phase and multi-component fluids and a
number of corresponding extensions of the method have been published in
the past \cite{Shan1993, bib:shan-chen:1994,
bib:orlandini-swift-yeomans:1995, bib:swift-orlandini-osborn-yeomans:1996,
bib:dupin-halliday-care:2003, bib:lishchuk-care-halliday:2003}.
In the Shan-Chen multi-component model \cite{Shan1993}, a system containing $N$ miscible or
immiscible fluids (with index $\sigma$) are described by $N$ sets of
distribution functions $f_{\sigma,i}$, one for each species. As
a consequence, $N$ lattice Boltzmann equations with relaxation times
$\tau_\sigma$ have to be considered. The interaction between the fluid
species is mediated via a local force density,
\begin{equation}
\label{eq:shanchen}
{\bf F}_{\sigma}({\bf r},t) = -\Psi_\sigma({\bf r}, t) \sum_{\sigma'} g_{\sigma \sigma'} \sum_{{\bf r}'} \Psi_{\sigma'}({\bf r}',t)({\bf r}' - {\bf r}).
\end{equation}
Here, the contributions of all species $\sigma'$ are taken into account,
and the force density acting on species $\sigma$ is obtained. The sum runs
over all neighbouring sites ${\bf r}'$ of site ${\bf r}$. The factor
$g_{\sigma \sigma'}$ is the coupling constant defining the interaction
strength between species $\sigma$ and $\sigma'$. It is related to the
surface tension between both species. Self-interactions ($\sigma' =
\sigma$) are also allowed. The pseudopotential $\Psi_\sigma$ is a function
of the density $\rho_\sigma$, in our case $\Psi_\sigma = 1 - \exp(- \rho_\sigma)$. Its shape is related to the equation of
state. There are different ways to incorporate forces such as those in
Eq.~(\ref{eq:shanchen}) into the LBM. One possibility is the so-called
velocity shift in the equilibrium distribution function: For the
computation of the equilibrium distribution $f_{\sigma,i}^{\text{eq}}$ of
component $\sigma$, ${\bf u}$ in Eq.~(\ref{eq:equilibrium}) is replaced by
$\left({\bf u}^{\text{eq}} + \tau_\sigma {\bf F}_\sigma /
\rho_\sigma\right)$, where $\rho_\sigma$ is the density of component
$\sigma$ and ${\bf u}^{\text{eq}} = \sum_\sigma \frac{\rho_\sigma {\bf
u}_\sigma}{\tau_\sigma} / \sum_\sigma \frac{\rho_\sigma}{\tau_\sigma}$ is
the common equilibrium velocity of all components. The flow velocity of
the fluid is then given by ${\bf u} = \sum_\sigma(\sum_i f_{\sigma,i} {\bf
e}_i + {\bf F}_\sigma/2) / \rho$. The Shan-Chen model belongs to the class
of front capturing methods. It is suitable to track interfaces for which
the topology evolves in time, for example, the breakup of a droplet.

In section \ref{sec:results}, we give examples where the LBM is used as a fluid flow solver to study the dynamics of complex interfaces.

%

\subsection{Continuum interface models}
\label{sec:macroscopic}
In this section, we present the \textit{immersed boundary method} (IBM) as a front tracking approach \cite{Peskin2002}. Such methods are mostly employed when the interface is formed by an additional continuous material whose constitutive behaviour (e.g., elasticity, viscosity) is assumed to be known. Examples are capsules, vesicles, or biological cells. It is usually not necessary to resolve the physical mechanisms which lead to this constitutive behaviour since there is a significant scale separation (if one is only interested in the mechanical properties on the macroscale). The IBM and related front-tracking approaches can also be used to describe the motion of the interface even when it is not formed by an additional material. For example, the interface between a gas and a liquid phase may be tracked \cite{tryggvason_front-tracking_2001}.

Within the IBM, the interface is considered sharp (zero thickness) and is represented by a cluster of marker points (nodes) which constitute a moving Lagrangian mesh (cf.\ Fig.~\ref{fig:lattice_etc}(a)). This mesh is immersed in a fixed Eulerian lattice representing the fluid. The question is how to predict the motion of the interface in time and how it affects the fluid. To consider correct dynamics, a bi-directional coupling of the lattice fluid and the moving Lagrangian mesh has to be taken into account. On the one hand, the interface is modelled as an impermeable structure obeying the no-slip condition at its surface. It is assumed that the flow field is continuous across the interface and that the interface is massless. Therefore, the interface is moving along with the ambient fluid velocity. On the other hand, a deformation of the interface generally leads to stresses reacting back onto the fluid via local forces. As an example, interfacial stresses are caused by local bending of a fluid-fluid interface (surface tension) or shearing of a capsule membrane (shear deformation). The stresses depend on the chosen constitutive behaviour of the interface and are not predicted within the IBM itself. This makes the IBM a flexible coupling scheme as it does not dictate specific material properties. The two-way coupling is accomplished in two main steps: (i) velocity interpolation and Lagrangian node advection and (ii) force spreading (reaction).

\textit{Interpolation}: First, the fluid flow field is computed such that the velocity $\bf u$ is known at each lattice site $\bf x$. In principle, any Navier-Stokes solver is applicable; in our case, we opted for the LBM, resulting in an immersed boundary-lattice Boltzmann method. Afterwards, the fluid velocity is interpolated to obtain its value at the position ${\bf r}_i$ of each membrane node $i$ which equals the velocity $\dot{\bf r}_i$ of that node:
\begin{equation}
 \dot{\bf r}_i = \sum_{\bf x} \Delta({\bf x} - {\bf r}_i) {\bf u}({\bf x}).
\end{equation}
The interpolation stencil $\Delta({\bf x} - {\bf r}_i)$ is a suitable discretised approximation of Dirac's delta function \cite{Peskin2002}. For numerical efficiency, its range should be finite. Peskin suggested several stencils with different interpolation ranges, e.g., taking into account two, three, or four lattice sites along each coordinate axis \cite{Peskin2002}. While the 2-point stencil is numerically faster and leads to a sharper description of the interface, the 3- and 4-point stencils result in smoother interpolated velocity fields \cite{kruger2011efficient}. Finally, the interface is advected by updating the position of each membrane node using an Euler scheme, ${\bf r}_i(t + \Delta t) = {\bf r}_i(t) +  \dot{\bf r}_i(t) \Delta t$.

\textit{Reaction}: By advecting the interface shape, it is generally deformed. The new shape of the interface is not necessarily its equilibrium shape which would minimise its energy. Therefore, each node exerts a reaction force ${\bf F}_i$ (which is computed from the known deformation state and the constitutive interface properties) on its surrounding fluid. This force is distributed (also called ``spread'') to the fluid using the same stencil as for the velocity interpolation,
\begin{equation}
{\bf f}({\bf x}) = \sum_{i} \Delta ({\bf x} - {\bf r}_i) {\bf F}_i.
\end{equation}
This force has to be taken into account by the Navier-Stokes solver as a local acceleration in the next time step.

\subsection{Mesoscopic interface models}
\label{sec:mesoscopic}

Surfactants have been treated numerically for many years and as such, the models and descriptions have evolved over time. Even though amphiphilic surfactant molecules derive many important properties from their structure at microscopic scales (the presence of molecular hydrophilic ``head'' and hydrophobic ``tail'' groups), modelling them as individual molecules is generally unfeasible, computationally, if one is interested in effects on macroscopic or mesoscopic scales. For example, in the case of surface tension reduction at a fluid-fluid interface, this would not supply any detail that is not obscured by the rheology; thus, surfactants are often modelled with some degree of coarse-graining. This makes this strategy well-suited to be coupled to other mesoscopic methods.

Simulations based on lattice-gas-automata (LGA) have been used since the nine\-ties. As a strongly simplified scheme, only the effect that the surfactant has on surface tension can be modelled~\cite{bib:chen-lookman:1995}. Gompper and Schick have introduced a model which allows the surfactant to have orientational degrees of freedom, as well as translational ones~\cite{bib:gompper-schick:1994}, which due to the shape and nature of the amphiphiles is a necessary property to simulate. Just as the LBM has evolved from LGA, the attendant surfactant extensions have found their way into LB as well. These implementations, built as an extension of LB, are --- unlike LGA-based models --- not suitable for simulating phenomena that strongly depend on effects caused by discrete particles. However, they are perfectly adequate in the area of bulk hydrodynamic phenomena which is of interest here. For example, the model developed by Chen et al.~\cite{bib:chen-boghosian-coveney-nekovee:2000, bib:nekovee-coveney-chen-boghosian:2000, bib:furtado-skartlien:2010} uses self-consistent forces to describe the interactions. This method introduces the surfactant as both an additional scalar field to model the local densities and a vector field to model local average dipole orientations (which can vary continuously). These fields are two-way coupled to the non-amphiphilic species in the system through Shan-Chen-type interaction forces. A similar force also describes a self-interaction to model the attraction between two amphiphile tails and repulsion between a head and a tail. This method is rather simple in its implementation, and as it assumes only interactions with neighbouring lattice sites, it is local and fast. However, the effective surface tension reduction that can be effected through this model is rather modest~\cite{frijters2012effects}.

Alternative methods built upon the LBM include the Ginzburg-Landau free energy-based model with two scalar order parameters introduced by Lamura et al.~\cite{bib:lamura-gonnella-yeomans:1999} and an extension to Chen's method which includes mid-range interactions by using more than one Brillouin zone in the calculations~\cite{bib:benzi-chibbaro-succi:2009, bib:benzi-sbragaglia-succi-bernaschi-chibbaro:2009}. This increases the range of surface tension reductions that can be achieved at the cost of reduced computational efficiency.

\subsection{Discrete interface models}
\label{sec:microscopic}

If there is no clear scale separation between immersed particles and the
lateral interface extension, it may be necessary to model the particles
explicitly. Here, we consider an ensemble of particles with well-defined
wetting behaviour. The particles themselves may have a complex shape and can be
provided with a constitutive model (of which the simplest is a rigid particle
model as we use it here).  There are several approaches to simulate a system of
two immiscible fluids and particles. One example which has recently been
applied by several groups is Molecular Dynamics (MD) coupled to the
LBM~\cite{jansen2011bijels,frijters2012effects,Guenther2012a,stratford2005colloidal,kim2010bijels,bib:joshi-sun:2010,bib:joshi-sun:2009}.
The advantage of this combination is the possibility to resolve the particles
as well as both fluids in such a way that all relevant hydrodynamical
properties are included. The particles are generally assumed to be rigid and
can have arbitrary shapes, where we restrict ourselves to spheres and
ellipsoids.

For the fluid-particle coupling, the particles are discretised on the lattice:
sites which are occupied by a particle are marked as solid. Following the
approach proposed by Ladd \cite{Ladd2001}, populations propagating from fluid
to particle sites are bounced back in the direction they came from (cf.\
Fig.~\ref{fig:lattice_etc}(b)). In this process, the populations receive
additional momentum due to the motion of the particles. In order to satisfy the
local conservation laws, linear and angular momentum contributions are assigned
to the corresponding particles as well. These in turn are used to update the
particle positions and orientations.  In the examples presented in this paper,
a leapfrog integrator is used to solve the equations for linear and angular
particle momenta.  In the context of the Shan-Chen multi-component algorithm
coupled to the particle solver, the outermost sites covered by a particle are
filled with a virtual fluid corresponding to a suitable average of the
surrounding unoccupied sites. This approach provides accurate dynamics of the
two-component fluid near the particle surface. The wetting properties of the
particle surface can be controlled by shifting the local density difference of
both fluid species by a given amount $\Delta \rho$. It can be shown that the
contact angle $\theta$ of the fluid-fluid interface at the particle surface has
a linear dependence on $\Delta\rho$~\cite{jansen2011bijels,Guenther2012a}.
Occasionally, when the particles move, lattice sites change from particle to
fluid state, which has to be treated properly
\cite{jansen2011bijels,frijters2012effects}.

If particles approach each other so closely that the lattice resolution is not
sufficient to resolve their hydrodynamic interaction, an additional lubrication
correction is required~\cite{Nguyen2002,jansen2011bijels,Guenther2012a}.
Further, a Hertzian potential is used to approximate an elastic interaction
between pairs of particles. For two overlapping spheres of radius $R$ with
mutual centre distance $r_{ij} < 2R$, it is defined as $\phi_{\text H} = K_{\text H} (2R - r_{ij})^{5/2}$ \cite{Hertz1881} where the force magnitude is controlled by
$K_{\text H}$. A generalisation for ellipsoids, based on the method of Berne and
Pechukas \cite{Berne1972}, is explained in \cite{Guenther2012a}.

\section{Case studies}
\label{sec:results}

\subsection{Continuum interface examples}
\label{sec:results_macro}

In this section, we show two exemplary applications of the immersed boundary-lattice Boltzmann method as introduced in section \ref{sec:macroscopic}. First, we present two-dimensional simulations of the dynamics of a viscous vesicle in simple shear flow. In section \ref{sec:platelets}, three-dimensional simulations of the separation of blood components are shown.

\subsubsection{Tumbling and tank-treading of viscous vesicles}

A vesicle is a closed phospholipid membrane filled with a fluid. Vesicles
can be used to micro-encapsulate active materials for drug delivery. They
also pose a biomimetic model for biological cells. One of the questions
that attracted the interest of scientists in the past decades was: how
does a vesicle behave dynamically under flow? There exists a rich
literature related to this problem (e.g., \cite{Keller1982, Mader2006,
Kantsler2006}). Today it is known that a vesicle undergoes either
tank-treading (steady motion) or tumbling (unsteady motion) when it is
subjected to simple shear flow. In the tank-treading mode, the vesicle's
main axis assumes a steady inclination angle $\theta$ with the flow
direction while its membrane undergoes a tank-treading like motion. In the
tumbling mode, the vesicle rotates as a solid, elongated particle.

Here, we reproduce these dynamical states via the combined immersed boundary-lattice Boltzmann method in two dimensions at negligible Reynolds numbers \cite{Kaoui2011a, Kaoui2012}. The vesicle encloses an internal fluid and is suspended in an external fluid. Both fluids are considered incompressible and Newtonian. In the present case, we are interested in viscous vesicles where the internal and external fluid viscosities differ in general. Their ratio $\Lambda := \frac{\eta_{\text{int}}}{\eta_{\text{ext}}}$ is called viscosity contrast. Numerically, we distinguish between the two fluids by the \textit{even-odd rule}, an algorithm deciding whether a fluid site is inside or outside the vesicle \cite{Kaoui2012}. Depending on the location, the local fluid viscosity is set to either $\eta_{\text{int}}$ or $\eta_{\text{ext}}$ via two different lattice Boltzmann relaxation times ($\tau_{\text{int}}$ and $\tau_{\text{ext}}$). The vesicle membrane (interface) influences the ambient fluids (inside and outside) by exerting a local force on them. This force results from the resistance of the membrane to bending and stretching/compression \cite{Kaoui2008}. It is computed on the membrane according to
\begin{equation}
 {\bf f} = \left[\kappa_{\text B} \left( \frac{\partial ^2 c}{\partial s ^2} + \frac{c^3}{2} \right) - c\, \tension \right]{\bf n} + \frac{\partial \tension}{\partial s}{\bf t},
\end{equation}
where ${\bf n}$ and ${\bf t}$ are the normal and tangential unit vectors, respectively. The membrane is characterised by the membrane bending modulus $\kappa_{\text B}$ and the local effective tension $\tension$. The local curvature is $c$, and $s$ is the arclength coordinate. The 4-point immersed boundary stencil is used to couple the dynamics of the membrane on the Lagrangian mesh and the fluid on the Eulerian lattice.


\begin{figure}
 \subfloat[]{\shortstack{\includegraphics[angle=90,width=0.13\linewidth]{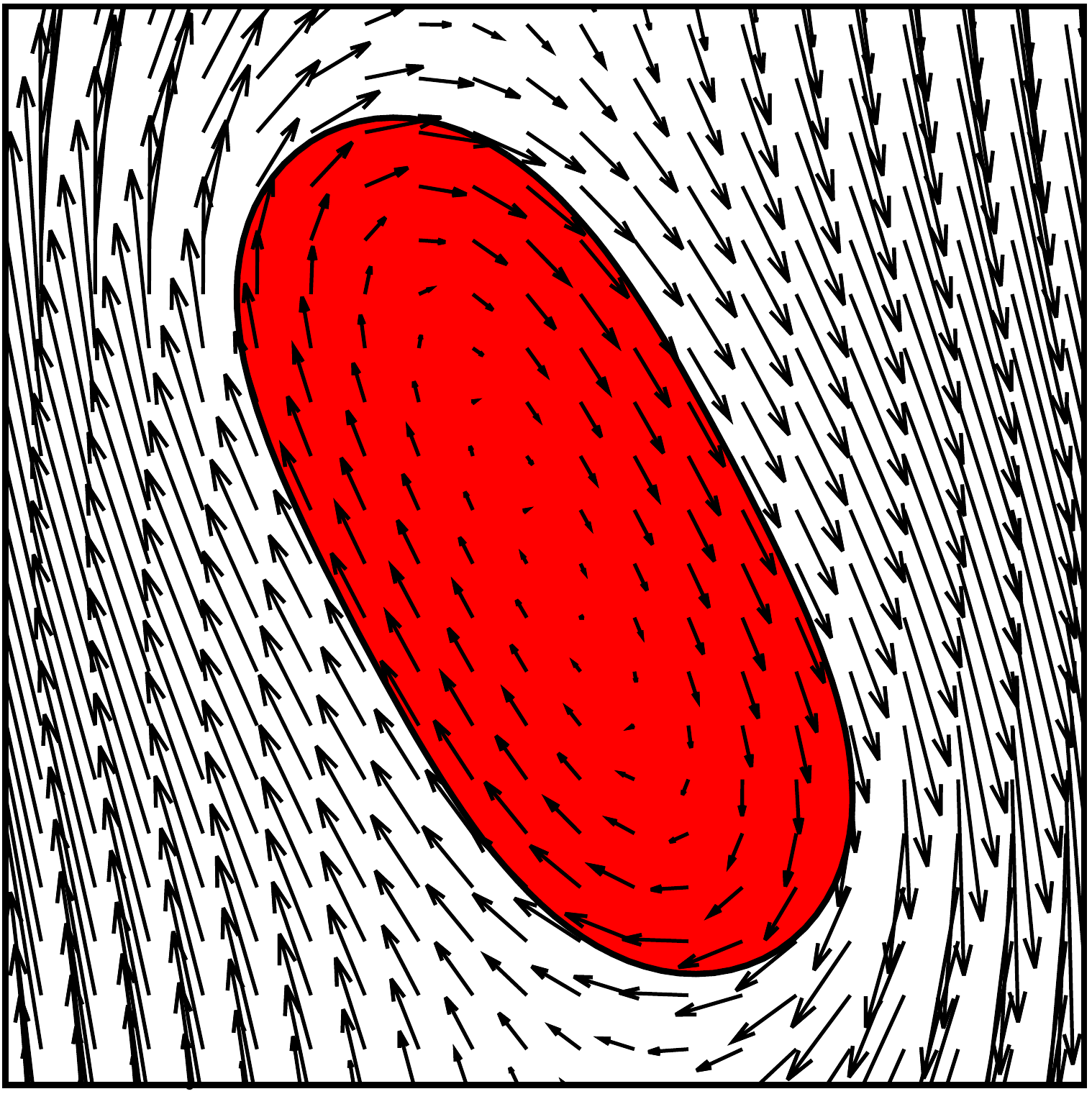}}}
 \qquad
 \subfloat[]{\shortstack{\includegraphics[angle=90,width=0.13\linewidth]{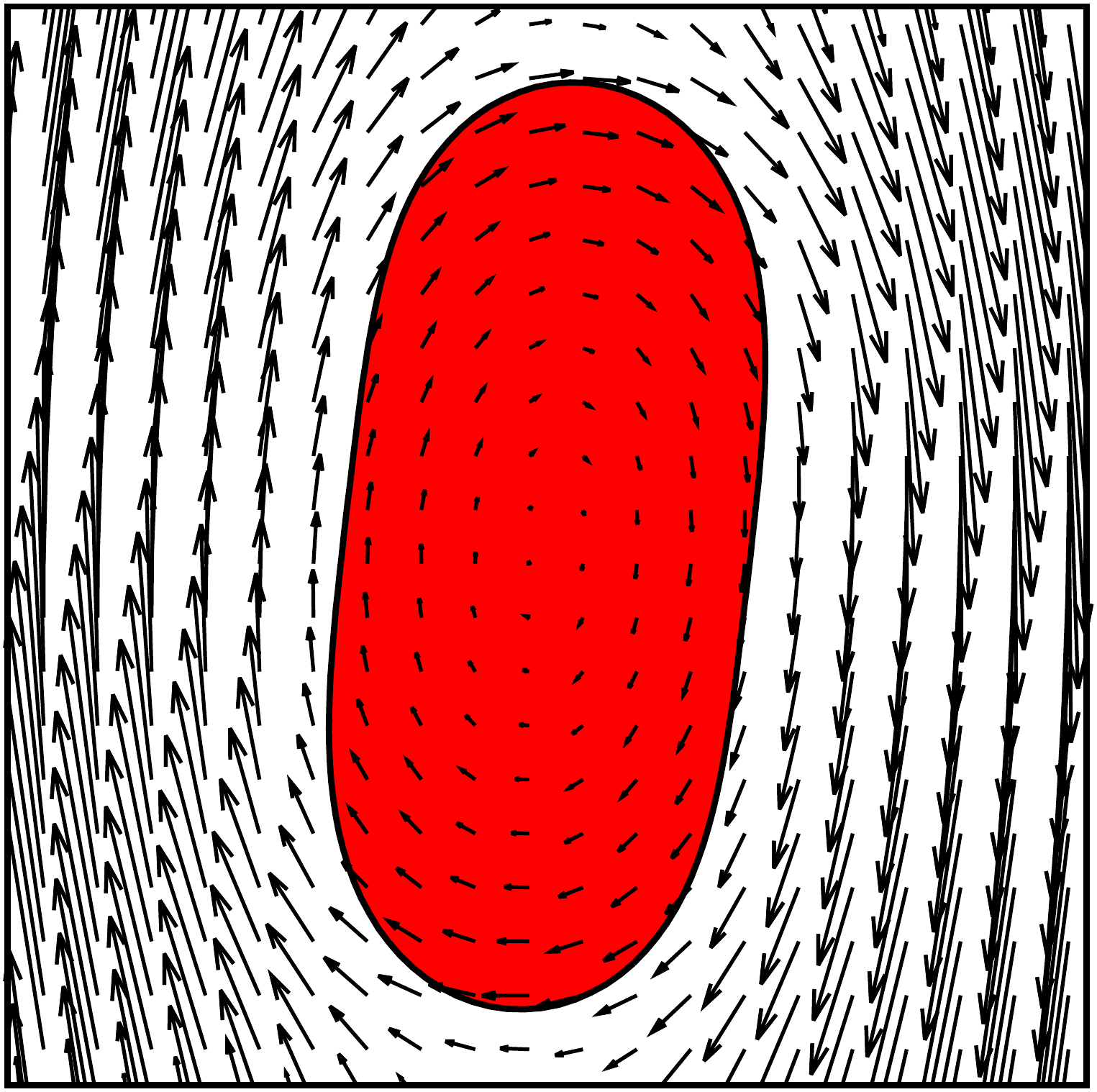} \includegraphics[angle=90,width=0.13\linewidth]{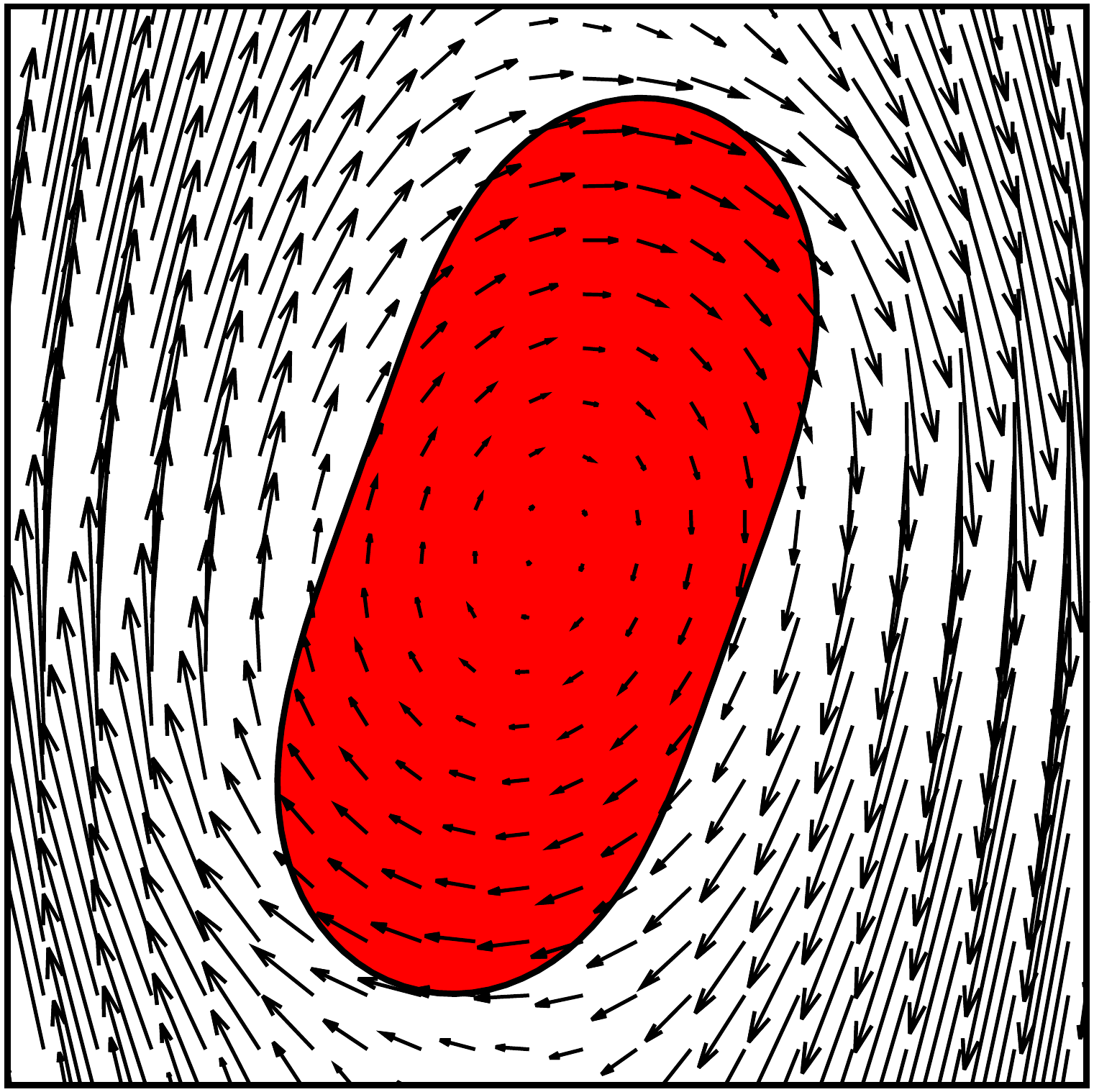} \includegraphics[angle=90,width=0.13\linewidth]{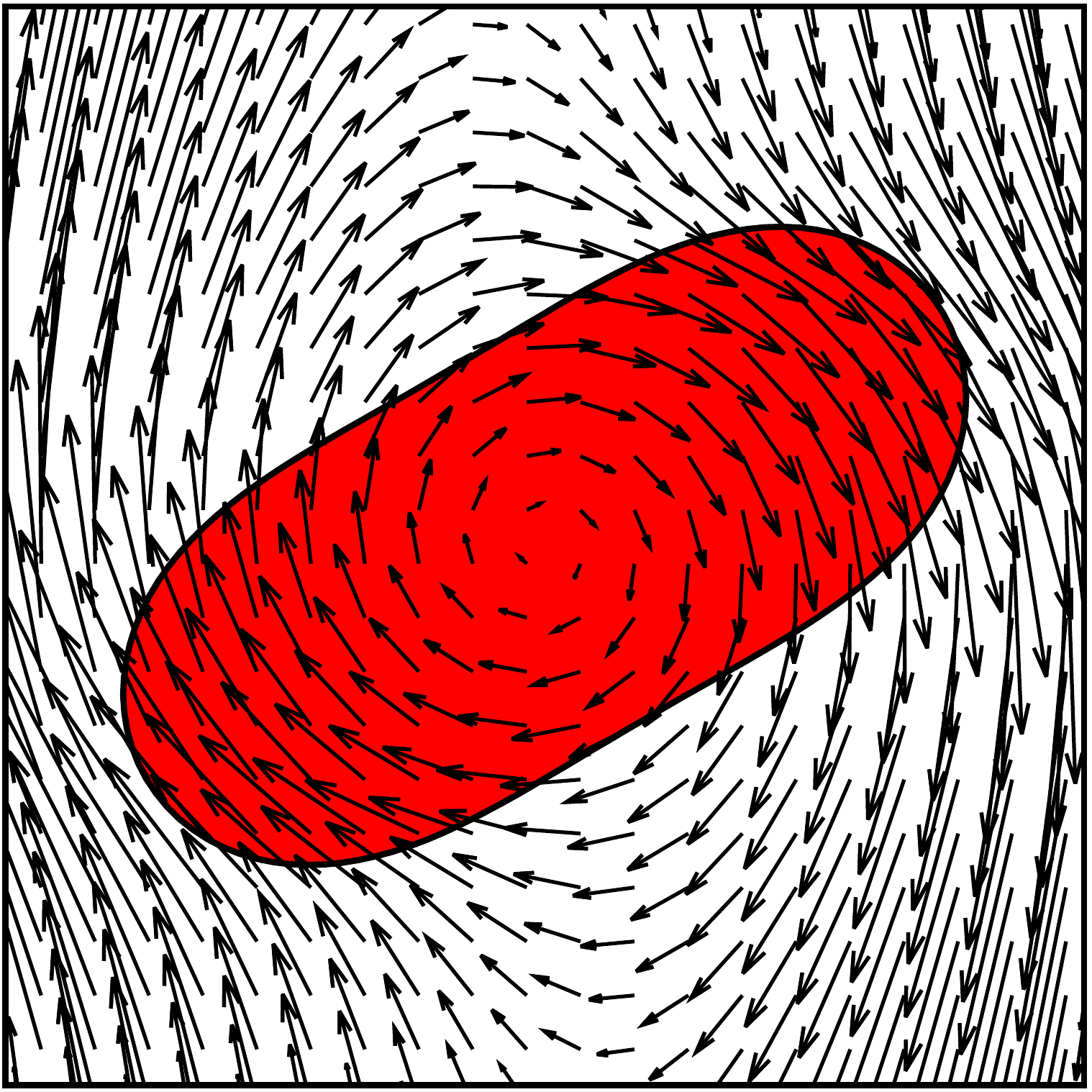}  \includegraphics[angle=90,width=0.13\linewidth]{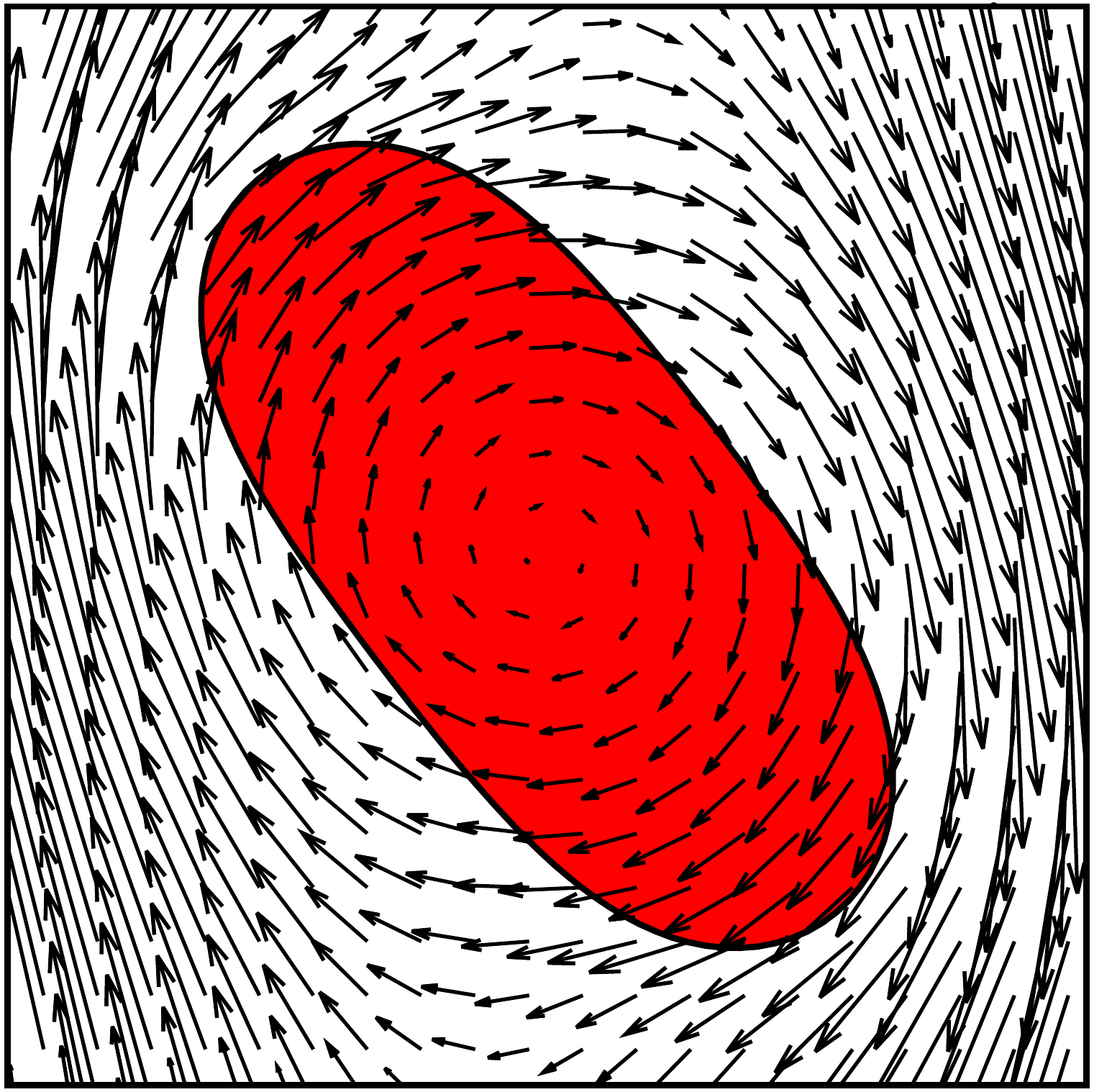} \includegraphics[angle=90,width=0.13\linewidth]{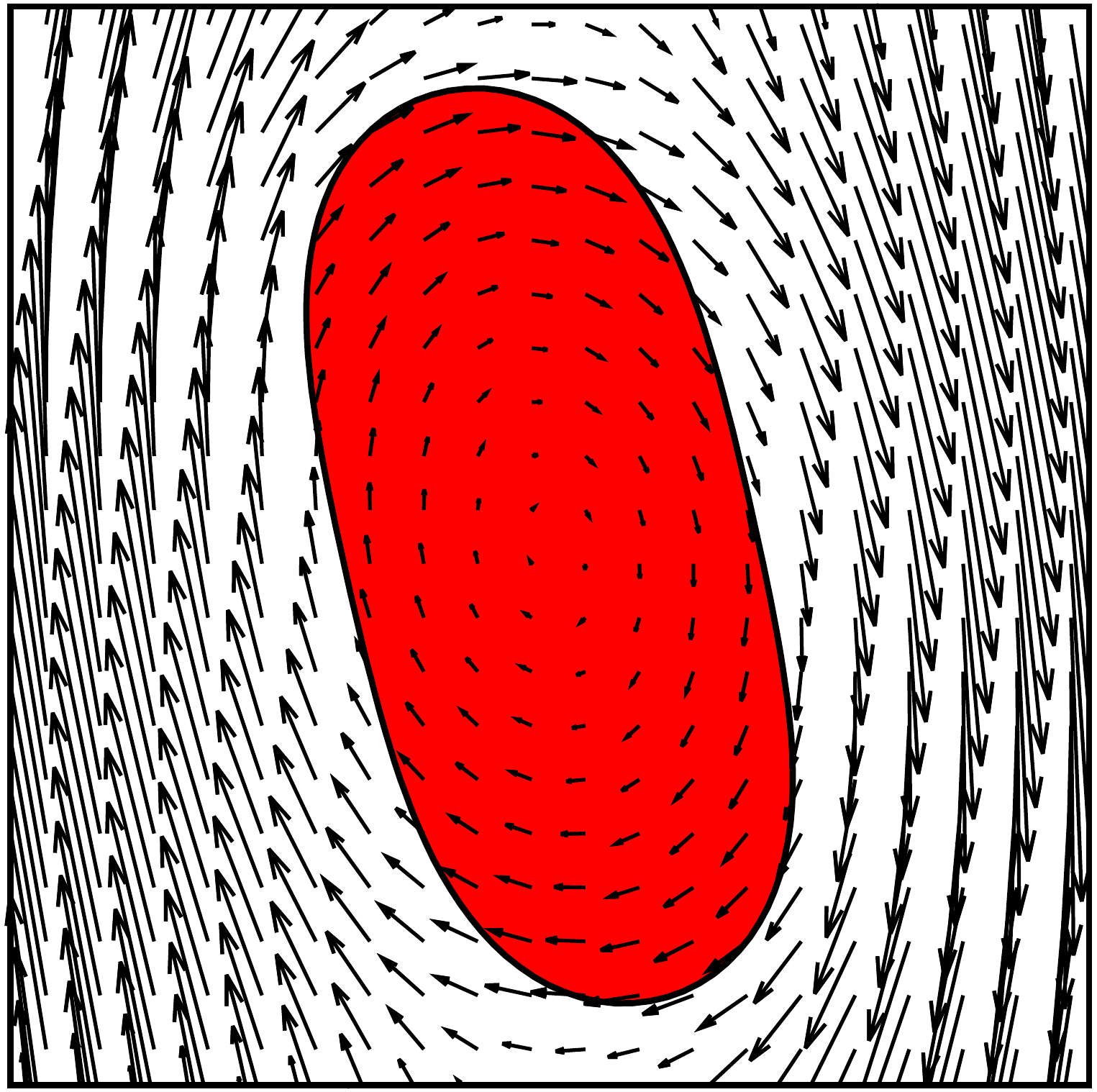} \includegraphics[angle=90,width=0.13\linewidth]{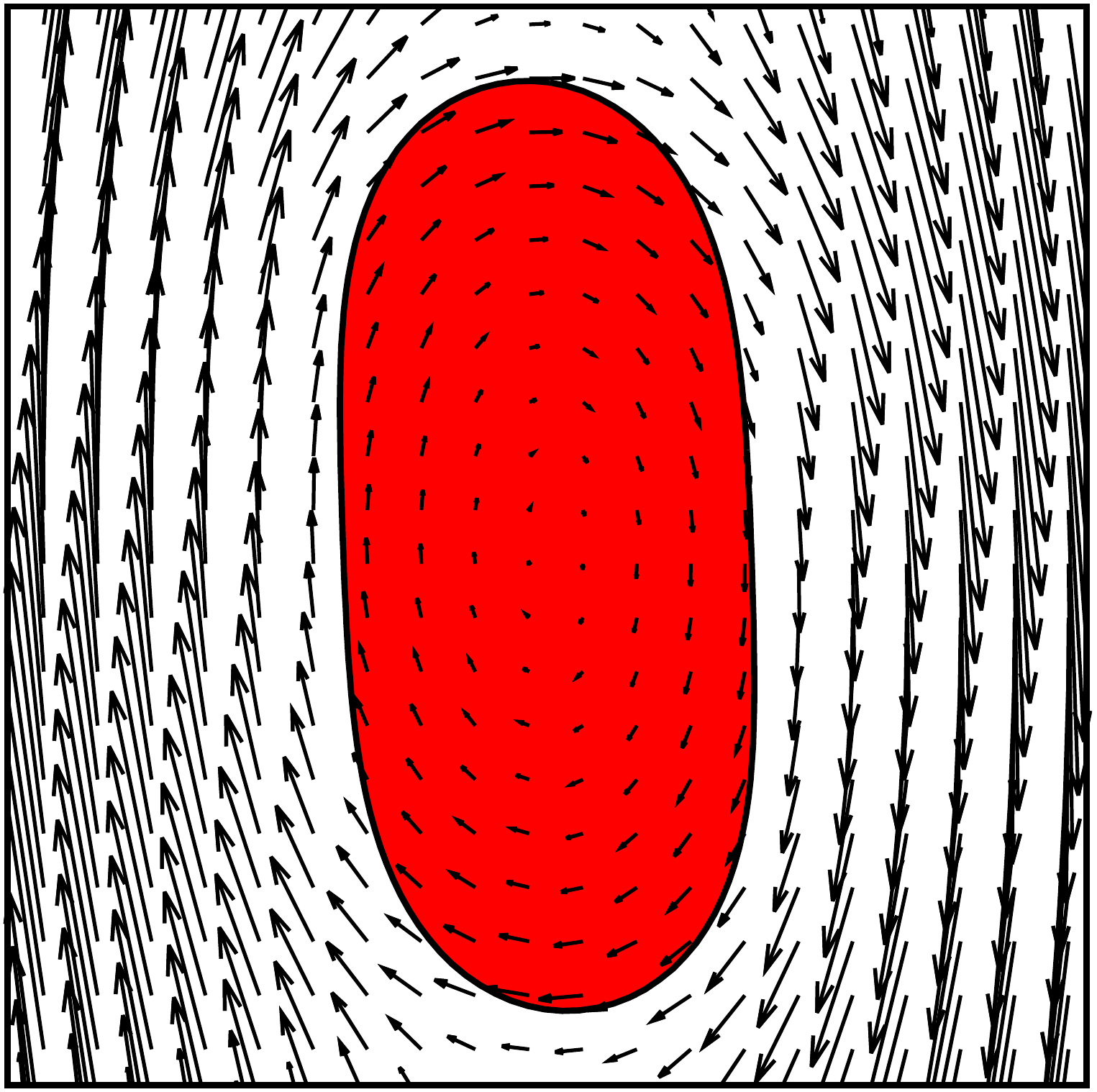}}}
 \caption{\label{fig:vesicle_snapshots} Snapshots showing the dynamics of a viscous vesicle under shear flow. (a) Tank-treading: the vesicle assumes a steady inclination angle with the flow direction while its membrane undergoes a tank-treading motion. (b) Tumbling: the vesicle rotates as a rigid elongated particle. The six snapshots (from left to right) illustrate the time evolution of the inclination angle, cf.\ Fig.~\ref{fig:vesicle_angles}.}
\end{figure}

\begin{figure}
 \centering
  \includegraphics[width=0.45\linewidth]{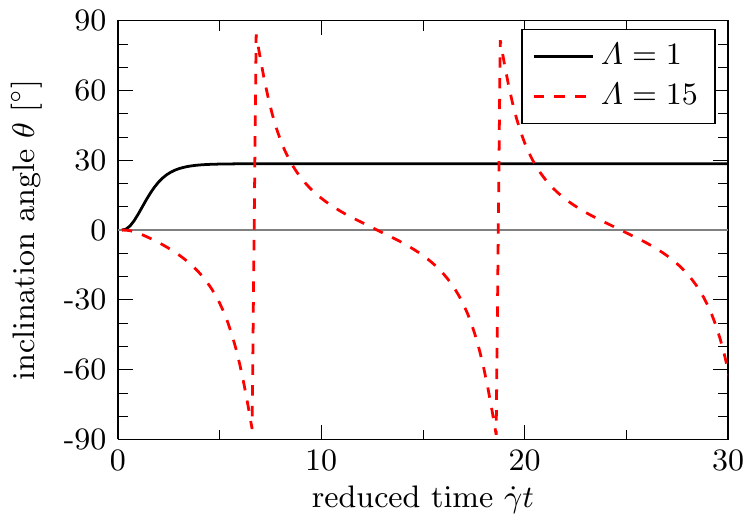}
 \caption{\label{fig:vesicle_angles} Time evolution of the vesicle inclination
 angle $\theta$ during the tank-treading (viscosity contrast $\Lambda = 1$) and tumbling ($\Lambda
 = 15$) modes. The time is scaled by the shear rate $\dot{\gamma}$. Some
 snapshots of the vesicle are shown in Fig.~\ref{fig:vesicle_snapshots}.}
\end{figure}

Fig.~\ref{fig:vesicle_snapshots} shows the dynamics of a vesicle in a simple shear flow with shear rate $\dot \gamma$. For a smaller viscosity contrast ($\Lambda = 1$), the vesicle tank-treads (Fig.~\ref{fig:vesicle_snapshots}(a)). The tank-treading motion of the vesicle membrane generates a rotational flow inside the vesicle. By increasing the viscosity contrast above a certain threshold, we induce a transition from tank-treading to tumbling. Fig.~\ref{fig:vesicle_snapshots}(b) shows the tumbling motion of a vesicle with $\Lambda = 15$. In Fig.~\ref{fig:vesicle_angles}, we report the time evolution of the inclination angle $\theta$ for both cases. Initially, the vesicle main axis is parallel to the flow axis ($\theta = 0$ at $t = 0$). In the tank-treading mode, $\theta$ increases until it assumes a steady value. Contrarily, in the tumbling mode, $\theta$ varies periodically.

It can be seen that the relatively simple immersed boundary-lattice Boltzmann method can be successfully employed to study vesicle dynamics in external flow fields. In particular, no remeshing of any kind is required.

\subsubsection{Flow-induced separation of red blood cells and platelets}
\label{sec:platelets}

Efficient separation of biological cells plays a major role in present-day medical applications, for example, the detection of diseased cells or the enrichment of rare cells. While active cell sorting relies on external forces acting on tagged particles, the idea of passive sorting is to take advantage of hydrodynamic effects in combination with the cells' membrane (interface) properties. It is known that deformable and rigid particles behave differently when exposed to external flow fields. For example, in Stokes flow ($\text{Re} = 0$), deformable capsules ($\text{Ca} > 0$) show a strong tendency to migrate towards the centreline of a pressure-gradient driven channel flow. Therefore, it has been proposed to use the particle deformability as intrinsic marker to separate rigid and deformable particles in specifically designed microfluidic devices. We show that the separation of red blood cells (RBCs) and platelets, which has been shown recently via experiments \cite{geislinger2012separation}, can be simulated by means of the LBM and immersed boundary method.

\begin{figure}
 \centering
 \subfloat[initial state]{\includegraphics[width=0.32\linewidth]{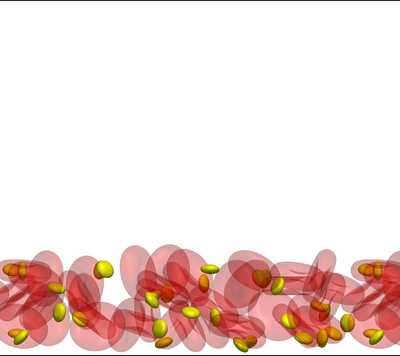}} \hfill
 \subfloat[final state, $\text{Ca} = 0.015$]{\includegraphics[width=0.32\linewidth]{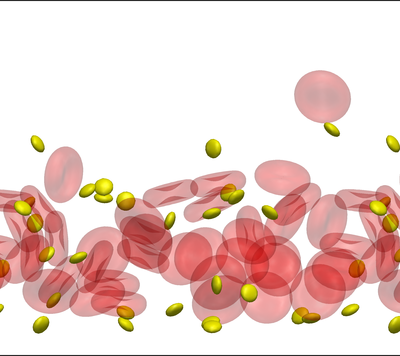}} \hfill
 \subfloat[final state, $\text{Ca} = 0.3$]{\includegraphics[width=0.32\linewidth]{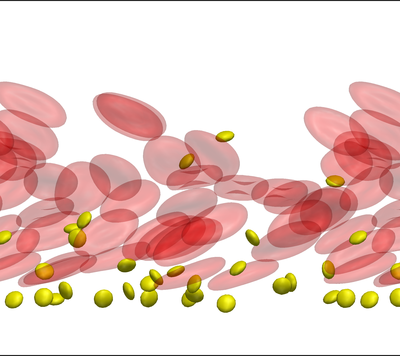}}
 \caption{\label{fig:casestudy_separation_1} Simulation snapshots of the
 separation of red blood cells (RBCs) and platelets. (a) Initially, the random
 suspension of 40 RBCs and 30 platelets is located in the bottom $30\%$ of the
 channel ($50\, \upmu\text{m}$ diameter, denoted by two black lines). RBCs
 (red) are shown with reduced opacity to reveal the platelet positions
 (yellow). Panels (b) and (c) show the lateral distribution of the cells after
 they have been moving downstream (rightwards) by about $20\, \text{mm}$ on
 average. For rigid and tumbling RBCs, lateral separation of RBCs and platelets
 is not pronounced (b). The separation is more efficient when the RBCs are
 strongly deformable and are tank-treading in the vicinity of the walls (c). The corresponding volume fraction profiles
 are shown in Fig.\ \ref{fig:casestudy_separation_2}.}
\end{figure}

The RBCs and platelets are modelled as closed two-dimensional membranes with identical fluids in the interior and exterior regions ($\Lambda = 1$). In the present approach, a finite element method is used to compute the local surface shear deformation and area dilatation and the resulting forces \cite{kruger2011efficient, kruger2011particlestress}. The 2-point stencil has been used for IBM interpolation and spreading. 40 RBCs and 30 platelets are randomly distributed in the bottom $30\%$ of a channel segment with $50 \times 50 \times 50\, \upmu\text{m}^3$ volume, cf.\ Fig.\ \ref{fig:casestudy_separation_1}. The platelets can be considered as rigid ellipsoidal particles. The large RBC and platelet diameters are $16$ and $5.5$ lattice constants, respectively. The channel is bounded by two parallel walls and is periodic in the other two directions. A force $f$ in rightward direction is used to drive the flow, resulting in a Poiseuille-like velocity profile. The average shear rate is defined as $\dot \gamma := 2 \hat u / H$ where $\hat u = f H^2 / (8 \eta)$ is the central velocity in the absence of particles, $H$ is the channel diameter, and $\eta$ is the external fluid viscosity. The RBC membrane (interface) is characterised by the capillary number $\text{Ca} = \eta \dot \gamma r / \kappa_{\text s}$. Here, $r = 4\, \upmu\text{m}$ is the large RBC radius and $\kappa_{\text s}$ is its shear elasticity which is about $5\, \upmu\text{N}\, \text{m}^{-1}$ for healthy RBCs. Two simulations have been run, one with the normal elasticity, leading to $\text{Ca} = 0.3$ for the selected viscosity and driving force. In the other case, the RBC shear modulus has been increased by a factor of 20, and the capillary number was $0.015$. This situation is typical for diseased RBCs (e.g., due to malaria or sickle cell anaemia).

\begin{figure}
 \centering
 \subfloat[initial state]{
  \includegraphics[height=4.5cm]{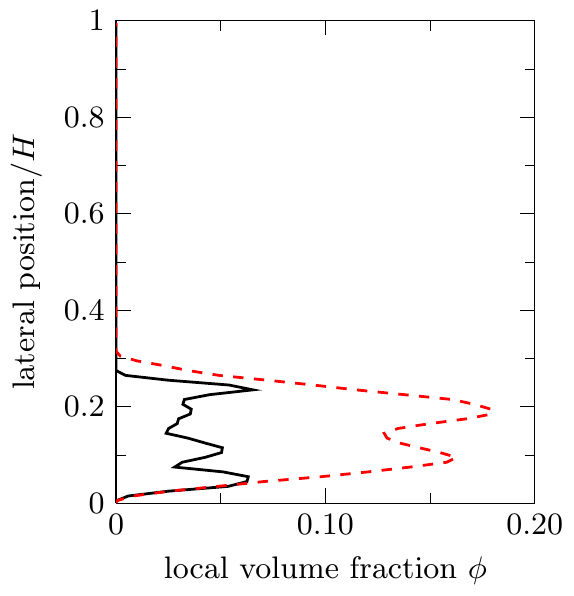}
 }
 \hfill
 \subfloat[final state, $\text{Ca} = 0.015$]{
  \includegraphics[height=4.5cm]{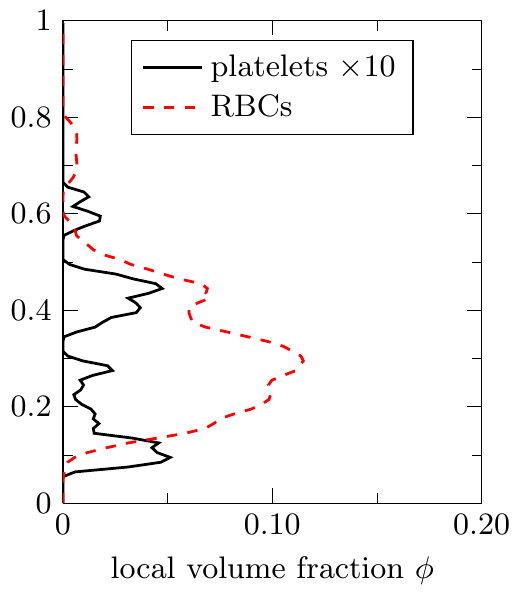}
 }
 \hfill
 \subfloat[final state, $\text{Ca} = 0.3$]{
  \includegraphics[height=4.5cm]{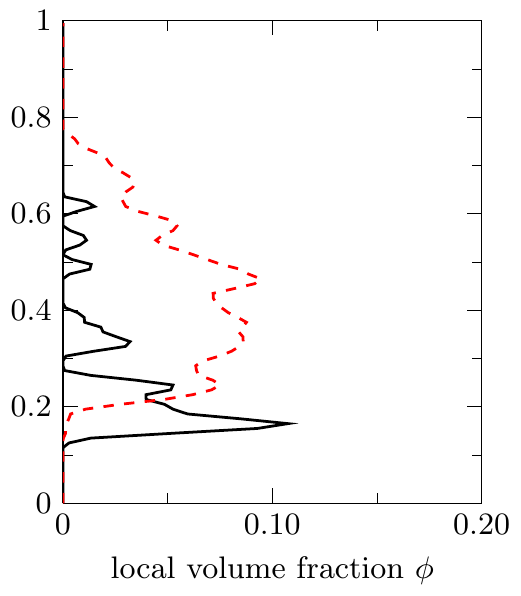}
 }
 \caption{\label{fig:casestudy_separation_2} Local volume fraction $\phi$
 (averaged over flow and vorticity directions) of red blood cells (RBCs, dashed
 red lines) and platelets (solid black lines) as function of lateral position.
 The two walls are located at $0$ and $H$. The platelet volume fractions are
 multiplied by a factor of 10 for better visibility. The plots are rotated by
 $90^\circ$ to enable a direct comparison of the profiles. (a) Initial state; state after about $20\, \text{mm}$ downstream motion for (b) rigid and (c) soft RBCs. Although platelet margination away from the RBC bulk can be seen in both cases, the platelet
 separation \textit{towards the bottom wall} is more efficient for softer RBCs. The three panels correspond to
 the snapshots in Fig.\ \ref{fig:casestudy_separation_1}.}
\end{figure}

The initial simulation states are shown in
Fig.~\ref{fig:casestudy_separation_1}(a) and
Fig.~\ref{fig:casestudy_separation_2}(a), while the corresponding panels (b)
and (c) show the state after about $20\, \text{mm}$ downstream motion for different capillary numbers (0.015 and 0.3). It
can be seen that RBCs and platelets can be efficiently separated when the RBCs
are sufficiently deformable (Fig.\ \ref{fig:casestudy_separation_1}(c) and
Fig.\ \ref{fig:casestudy_separation_2}(c)). The platelets marginate into the
gap forming between the bottom wall and the RBC bulk. Contrarily, for the rigid
RBCs, separation is visible but not pronounced (Fig.\
\ref{fig:casestudy_separation_1}(b) and Fig.\
\ref{fig:casestudy_separation_2}(b)). Additionally, the bulk of the RBCs moves
faster towards the centreplane when the particles are more deformable, i.e.,
when $\text{Ca}$ is larger. For $\text{Ca} = 0.30$, RBCs near the wall are
observed to tank-tread; for $\text{Ca} = 0.015$, all are tumbling. The lateral
migration timescale is several orders of magnitude larger than the timescale
for advection along the channel: the bulk of the soft RBCs has moved by only
$15\, \upmu\text{m}$ towards the centreplane after travelling about $20\,
\text{mm}$ in downstream direction.

This case study is a striking example for the effect of interface properties on the overall flow behaviour of suspensions. It also shows that the immersed boundary-lattice Boltzmann method can be applied to suspensions of soft particles at moderate volume fractions.

\subsection{Mesoscopic interface examples}
\label{sec:results_meso}

A mesoscopic description of interfaces stabilised by amphiphilic molecules
involves a systematic coarse-graining of the molecular degrees of freedom,
while keeping the effective mediated interactions between other fluids in the
system. This can be obtained using the model described in
section~\ref{sec:mesoscopic}, where the surfactant is simulated as a fluid phase
together with a local mean-field dipolar force mediating the amphiphilic
interactions on a mesoscopic level.  The applicability of the method is being
demonstrated in this section by two examples: first, we show how phase
separation and emulsion stabilisation can be simulated using a ternary LB
model. Second, the model is applied to study a self-organised amphiphilic
mesophase, namely the gyroid mesophase.

\subsubsection{Phase separation and emulsification of amphiphilic mixtures}
In section~\ref{sec:mesoscopic}, several mesoscopic methods to simulate amphiphilic
fluids were mentioned. Such fluids contain at least one species made of
surfactant molecules and are in general complex fluids. By adjusting
temperature, fluid composition or pressure, the amphiphiles can self-assemble
and force the fluid mixture into a number of equilibrium structures or
mesophases. These include lamellae and hexagonally packed cylinders, micellar,
primitive, diamond, or gyroid cubic mesophases as well as sponge phases~\cite{saksena-coveney-2008}. The
latter are generally not well ordered, but might have a well defined average size of
fluid domains. In the context of our simulations they are called bicontinuous
microemulsions since they are formed by the amphiphilic stabilisation of a
phase-separating binary mixture, where the two immiscible fluid constituents occur
in equal proportions. Here, the oil and water phases interpenetrate and
percolate and are separated by a monolayer of surfactant at the interface.

It has been shown by Langevin, molecular dynamics, lattice gas, and
lattice Boltzmann simulations that the temporal growth law for the size of
oil and water domains in a system without amphiphiles follows a power law
$t^\gamma$, with $\gamma$ being between $1/3$ and
$1$~\cite{bib:gonzalez-nekovee-coveney}.  Adding amphiphiles to a binary
mixture of otherwise immiscible fluids which is undergoing phase
separation can cause the separation to slow down.  The domain growth
behaviour then crosses over to a logarithmic growth $(\ln t)^\theta$. A
further increase of the surfactant concentration can lead to growth which
is well described by a stretched exponential form $A-B\exp(-C t^\delta)$.
Capital and Greek letters denote fitting parameters and $t$ is the
time~\cite{bib:jens-giupponi-coveney:2006,bib:emerton-coveney-boghosian,bib:gonzalez-coveney-2}.

\begin{figure}
\centerline{\includegraphics[width=0.9\linewidth]{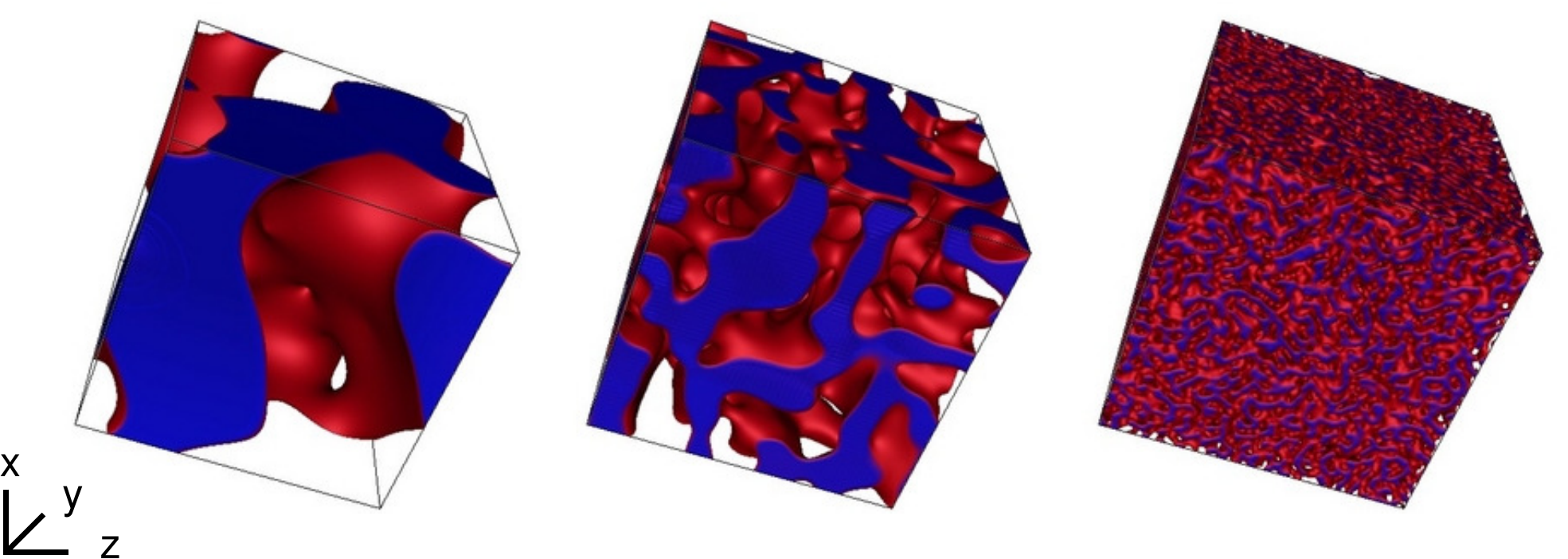}}
\caption{\label{fig:jens-Fig1} Volume rendered fluid densities for surfactant
densities $\rho_{\text s}$=0.00, 0.15, 0.30 (from left to right). Only one of the
immiscible fluid species is shown and different colours denote the interface and
areas of high density of the visualised fluid. The surfactant (not shown) is
mostly aligned at the interfaces and the second immiscible fluid component
fills the void space. After $3 \cdot 10^4$ time steps the phases have separated to a
large extent if no surfactant is present (left). Adding a small amount of
surfactant (centre) causes the domain growth to slow down.  For
high concentrations (right) the growth process arrests and a
stable bicontinuous microemulsion is formed
(from~\cite{bib:jens-giupponi-coveney:2007}).}
\end{figure}

In the simulations shown here we model two immiscible LB fluids and one
surfactant species. We avoid effects due to variations of the fluid densities
by keeping the total density in the system constant at 1.6 (in lattice units),
while varying the surfactant densities $\rho_{\text s}$ between 0.00 and 0.70.  To
depict the influence of the surfactant density on the phase separation process,
Fig.~\ref{fig:jens-Fig1} shows three volume rendered 256$^3$ systems at
surfactant densities 0.00 (left), 0.15 (centre), and 0.30 (right). After $3 \cdot 10^4$
time steps the phases are almost separated if no surfactant is present (left).
Longer simulation times would result in two perfectly separated phases. If one
adds some surfactant ($\rho_{\text s}=0.15$, centre), the domains grow more slowly,
visualised by the smaller structures in the volume rendered image. For
sufficiently high amphiphile concentrations ($\rho_{\text s}=0.30$, right) the growth
process arrests leading to a stable bicontinuous microemulsion with small
individual domains formed by the two immiscible fluids.

To investigate the influence of surfactant more quantitatively, we define the
time dependent lateral domain size $L(t)$ along direction $i=x,y,z$ as
\begin{equation}
 \label{eq:domsize}
 L(t)\equiv \frac{1}{3}\sum_{i=1}^3\frac{2\pi}{\sqrt{\left<k^2_i(t)\right>}},
\end{equation}
where $\left<k^2_i(t)\right>\equiv \left({\sum_\mathbf{k} k_i^2 S(\mathbf{k},t)}\right)/\left( {\sum_\mathbf{k} S(\mathbf{k},t)}\right)$
is the second order moment of the three-dimensional structure function $S(\mathbf{k},t)\equiv\frac{1}{V}\left|\phi^\prime_\mathbf{k}(t)\right|^2$
with respect to the Cartesian component $i$, $\left< \right>$ denotes the
average in Fourier space, weighted by $S(\mathbf{k}, t)$ and $V$ is the number
of sites of the lattice, $\phi^\prime_\mathbf{k}(t)$ the Fourier transform of
the fluctuations of the order parameter $\phi^\prime\equiv\phi-\left<\phi\right>$, and $k_i$ is the $i$th component of the wave vector.

\begin{figure}
 \centering
 \includegraphics[width=0.6\linewidth]{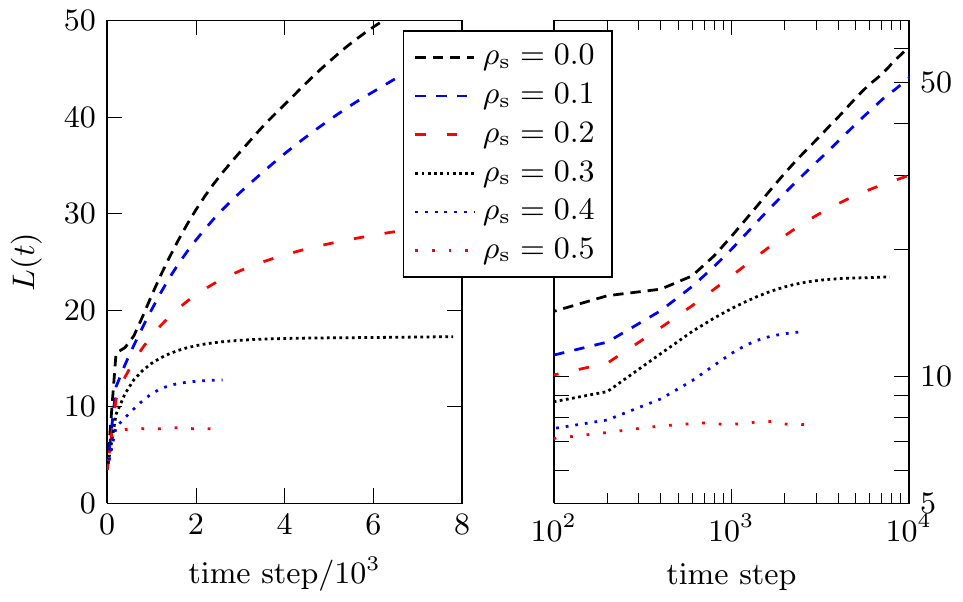}
 \caption{\label{fig:jens-Fig3} Linear (a) and logarithmic (b)
representation of the average domain size $L(t)$ for various surfactant
densities $\rho_{\text s}$ (data reprinted
from~\cite{bib:jens-giupponi-coveney:2007}). }

\end{figure}

In Fig.~\ref{fig:jens-Fig3} the time dependent lateral domain size $L(t)$ is
shown for a number of surfactant densities $\rho_{\text s}$ between 0.00 and 0.50.
Fig.~\ref{fig:jens-Fig3}(a) and (b) show identical data, but different scalings
of the axes. In (a), we plot the data linearly in order to give a better
impression of the time dependence of the growth dynamics. However, to determine
the actual growth law, we provide a log-scale plot of the same data in
Fig.~\ref{fig:jens-Fig3}(b). For the first few hundred time steps, the randomly
distributed fluid densities of the initial system configuration cause a
spontaneous formation of small domains resulting in a steep
increase of $L(t)$. For $\rho_{\text s} = 0.00$, domain growth does not come to an
end until the domains span the full system. By adding surfactant we can slow
down the growth process and for high surfactant densities, $\rho_{\text s} > 0.25$, the
domain growth stops after only a few thousand simulation time steps.  By adding
even more surfactant, the final average domain size decreases further.  We fit
our numerical data with the corresponding growth laws and find that for
$\rho_{\text s}$ smaller than 0.15 $L(t)$ is best fit by a function proportional to
$t^\alpha$. For $\rho_{\text s}$ being 0.15 or 0.20, a logarithmic behaviour
proportional to $(\ln t)^\theta$ is observed. Increasing $\rho_{\text s}$ further
results in $L(t)$ being best described by a stretched exponential. These
results correspond well with the findings in~\cite{bib:gonzalez-coveney-2}.

\subsubsection{The cubic gyroid mesophase}

\begin{figure}
\centerline{\includegraphics[width=0.5\linewidth]{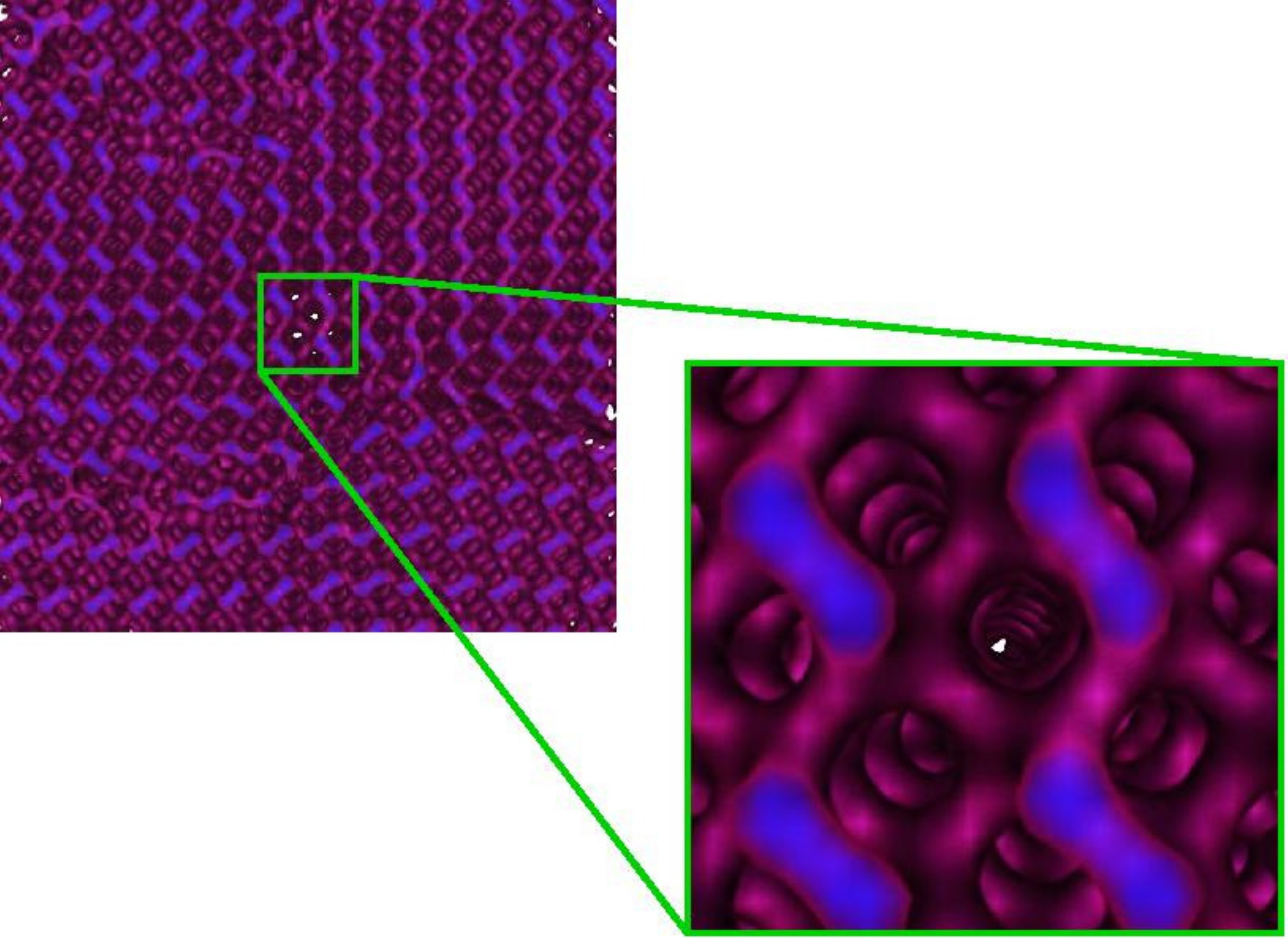}}
\caption{\label{fig:jens-gyroid}A volume rendered dataset from a ternary
fluid mixture forming a gyroid mesophase (from
\cite{bib:jens-harvey-chin-venturoli-coveney:2005}).}
\end{figure}

The ternary amphiphilic lattice Boltzmann method used here has been very
successfully applied to study the dynamical self-assembly of a particular amphiphile
mesophase, the gyroid
\cite{bib:gonzalez-coveney,bib:gonzalez-coveney-2,bib:jens-giupponi-coveney:2006,bib:jens-harvey-chin-venturoli-coveney:2005,bib:jens-harvey-chin-coveney:2004}.
This mesophase is observed to form from a homogeneous mixture of fluids,
without any external constraints imposed to let the gyroid geometry emerge.
This geometry is an effect of the mesoscopic fluid parameters only, which is
remarkable, since it allows insight into the dynamics of the mesophase
formation --- in contrast to most other treatments to date which have focused
on static properties or a mathematical description of the static equilibrium
state~\cite{bib:gandy-klinowski,bib:grosse-brauckmann}. In addition to its
biological importance, there have been recent attempts to use self-assembling
gyroids to construct nanoporous materials~\cite{bib:chan-hoffman-etal}.  The
gyroid structure can be understood as two labyrinths mainly consisting of
water and oil counterparts which are enclosed by the gyroid minimal surface at
which the surfactant molecules accumulate. Fig. \ref{fig:jens-gyroid} shows a
snapshot from a typical simulation of a gyroid mesophase and depicts the
characteristic triple junctions defined by the volumes containing the majority
of one of the individual fluid species. Multiple gyroid domains have formed and
the close-up shows the extremely regular, crystalline, gyroid structure within
a domain.

In small-scale simulations, the gyroid mesophase will evolve to perfectly fill
the simulated region, without defects. However, this can be accounted to finite
size effects. As the lattice size is being increased, numerous small, well
separated gyroid-phase regions or domains may start to form during the gyroid
self-assembly. While these domains grow independently, they will in general not
be identical, and might differ in orientation, position, or unit cell size.
Thus, grain-boundary defects arise between gyroid domains. Inside a domain,
dislocations, or line defects might occur, corresponding to the termination of
a plane of unit cells. There may also be localised non-gyroid regions,
corresponding to defects due to contamination or inhomogeneities in the initial
conditions.  Even with the longest possible simulation times, it is impossible
to generate a `perfect' crystal. Instead, either differently orientated domains
can still be found or individual defects are still diffusing around.
Understanding such defects is important for the comprehension of how these
phases behave dynamically and how one could produce and utilise them
experimentally~\cite{bib:jens-harvey-chin-coveney:2004}.

Of further interest is the rheological behaviour of the gyroid mesophase. It
has been shown that non-trivial rheological properties emerge when the gyroid
phase is placed in an oscillatory shear flow with frequency $\omega/2\pi$.
These include shear-thinning and even linear viscoelastic effects.
Furthermore, the ternary amphiphilic LB model correctly predicts the
theoretical limits for the moduli $G^{\prime}(\omega)$,
$G^{\prime\prime}(\omega)$ as $\omega$ goes to $0$ as well as a crossover in
$G^{\prime}(\omega)$, $G^{\prime\prime}(\omega)$ at higher
$\omega$~\cite{bib:jens-giupponi-coveney:2006,saksena-coveney-2009}.
It is remarkable that the model does not require any assumptions
at the macroscopic level in order to predict the formation and rheological
behaviour of cubic mesophases. Instead, it provides a purely kinetic approach
to the description of complex fluids.

\subsection{Discrete interface examples}
\label{sec:results_micro}

We now consider the effect of massive nanoparticles adsorped to a fluid-fluid
interface as an example of the coupled lattice Boltzmann-molecular dynamics
method explained in section \ref{sec:microscopic}. The first example shows the
ordering of ellipsoidal particles at the interface of a spherical fluid
droplet and compares it with the corresponding results of particles at a flat interface. This is followed by a case study highlighting how the presence of spherical nanoparticles affects the properties of a droplet in shear flow.

\subsubsection{Ordering of ellipsoidal particles at a fluid-fluid interface}

\begin{figure}
 \centering
 \subfloat[$t = 10^3$, $\chi = 0.153$]{\includegraphics[width=0.3\linewidth]{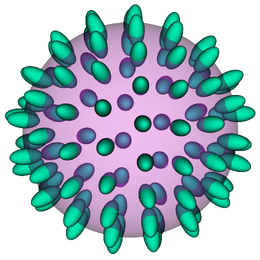}} \hfill
 \subfloat[$t = 10^3$, $\chi = 0.305$]{\includegraphics[width=0.3\linewidth]{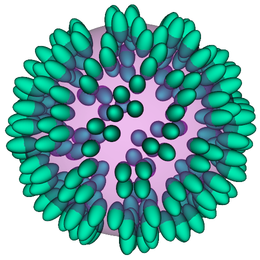}} \hfill
 \subfloat[$t = 2 \cdot 10^4$, $\chi = 0.305$]{\includegraphics[width=0.3\linewidth]{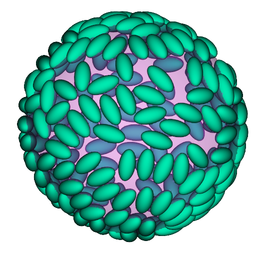}} \\
 \caption{\label{fig:droplet_snapshots} Snapshots of the particle-laden droplet
 for different times (in time steps) and coverage fractions $\chi$. One
 observes a higher global spatial order for the more dilute system in (a) as compared
 to the denser case in (b).
}
\end{figure}

Particles adsorbed at a fluid-fluid interface can stabilise emulsions of
immiscible fluids (e.g., Pickering emulsions where droplets of one fluid are
immersed in another fluid and are stabilised by colloidal particles)
\cite{jansen2011bijels,Guenther2012a}. These colloids are not necessarily
spherical such as a clay particle which has a flat shape. A simple
approximation of such a particle shape is an ellipsoid. The adsorption of a
single ellipsoid at a flat interface is the first step to understand the
dynamics of such emulsions \cite{Guenther2012a}. The next step is the study of
the behaviour of many ellipsoidal particles at an interface. In this section,
the behaviour of an ensemble of elongated ellipsoidal particles with aspect
ratio $m = 2$ at a single droplet interface is discussed \cite{Kaoui2011c}. The
results are compared to the corresponding case of a flat interface. The initial
particle orientation is such that the long axis is orthogonal to the local
interface (cf.\ Fig.\ \ref{fig:droplet_snapshots}(a)). As discussed in
\cite{Guenther2012a}, for the case of a single prolate ellipsoidal particle at
a flat interface, the equilibrium configuration is a particle orientation
parallel to the interface. We vary the coverage fraction
$\chi = N A_p / A_d$ which is defined as the ratio of the initial
interface area covered by the $N$ particles orientated orthogonal to the
interface and the total surface area of the spherical droplet $A_d$. $A_p$ is
the area which is covered by a single particle. Note that the covered area
increases when the particles rotate towards the interface. The simulations
discussed in this section have been performed for $\chi = 0.153$ and $\chi =
0.305$.

\begin{figure}
 \centering
 \subfloat[order parameter: \newline time evolution]{\includegraphics[width=0.325\linewidth]{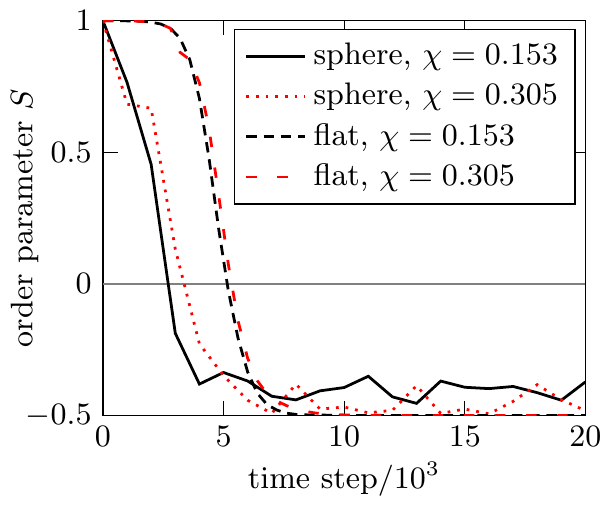}} \hfill
 \subfloat[pair correlation: \newline time dependence]{\includegraphics[width=0.325\linewidth]{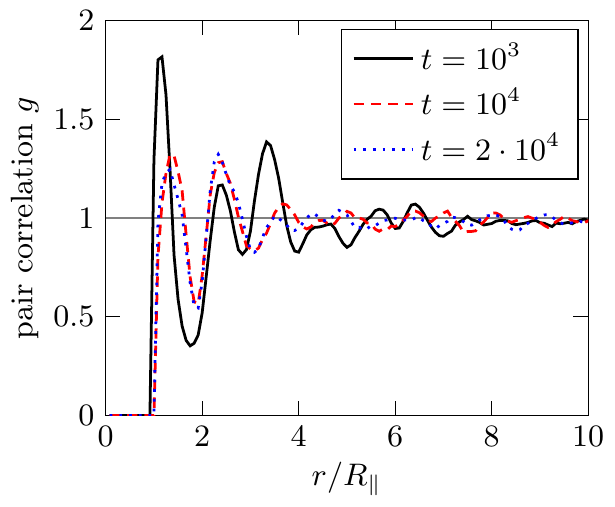}} \hfill
 \subfloat[pair correlation: \newline coverage dependance]{\includegraphics[width=0.325\linewidth]{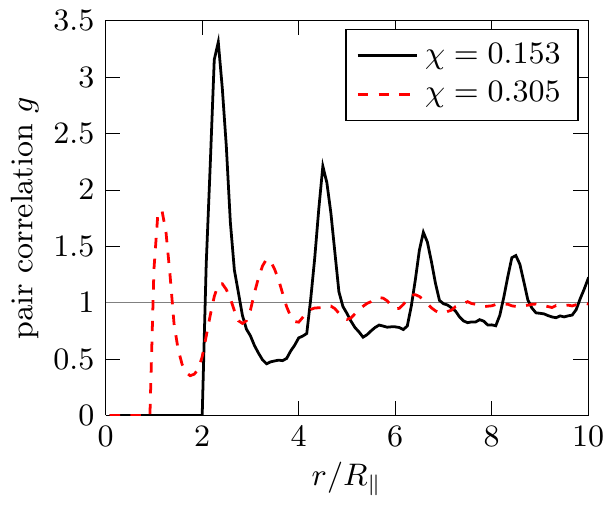}}

 \caption{\label{fig:SandGr} (a) Time evolution of the uniaxial order
   parameter $S$ (defined in Eq.~(\ref{eq:S})): the order decreases faster for
   the case of a sphere while the coverage fraction does not play an
   appreciable role. (b) The time dependence of the pair correlation function
   $g(r)$ (defined in Eq.~(\ref{eq:gd})) for particles adsorbed at a spherical
   interface ($\chi = 0.305$) is shown for different times. (c) Pair
   correlation function for the spherical interface after $10^3$ time steps
   for different values of $\chi$ corresponding to
   Fig.~\ref{fig:droplet_snapshots}(a) and (b). One can see that the order for
   the denser case is reduced due to capillary interactions which cause the
   particles to cluster causing a reduced long range order.
   }
\end{figure}

Fig. \ref{fig:droplet_snapshots}(a) and (b) show the particle laden droplet
almost at the beginning of the simulation (after $10^3$ time steps) where all
particles are perpendicular to the interface for both values of $\chi$. For the
case of $\chi = 0.305$, the particles are comparably close to each other so that
capillary interactions lead to the clustering motion of particles. The
reason for these capillary interactions is a deformation of the interface
caused by particles rotating towards the interface. In addition the rotating
particles cause a small deviation of the droplet from its originally spherical
shape which also leads to capillary interactions~\cite{Zeng2012a}.  In Fig.
\ref{fig:droplet_snapshots}(c), at the end of the simulation, the particles
have completed a rotation towards the interface and stabilise the largest
interfacial area possible. However, their local dynamics and temporal
re-ordering is a continuous interplay between capillary interactions,
hydrodynamic waves due to interface deformations and particle collisions.
Inspired from liquid crystal analysis, we use the uniaxial order parameter $S$
to characterise the orientational ordering of anisotropic particles,
\begin{equation}
 \label{eq:S}
  S=\frac{1}{2} \left\langle  3\cos^2\tilde{\vartheta}-1 \right\rangle
  \mbox{,}
\end{equation}
where $\tilde{\vartheta}$ is the angle between the particle main axis and the
normal of the interface at the point of the particle centre. The angular
brackets denote an instantaneous ensemble average. An ideal spherical shape is
assumed for the calculation of $S$. Fig. \ref{fig:SandGr}(a) shows the time
evolution of the order parameter $S$ for particles at a droplet interface for
$\chi = 0.153$ and $\chi = 0.305$ and the corresponding cases for particles at
a flat interface. Initially, the order parameter assumes the value $S_{\perp{}}
= 1$ which corresponds to perfect alignment perpendicularly to the interface.
In the case of flat interfaces, the order parameter reaches the final value of
$S\approx S_{\parallel{}} = -0.5$. This corresponds to the state where all
particles are aligned with the interface plane. For particle laden droplets
(cf.\ Fig.~\ref{fig:droplet_snapshots}(c)), a value of $S \approx -0.4 >
S_{\parallel{}}$ is found, which indicates that the particles are not entirely
aligned with the interface. Furthermore, $S$ fluctuates in time. One possible
reason for this deviation of $S$ from $S_\|$ is that the droplet does not have
the exact spherical shape which is assumed for the calculation of $S$. This
would lead to an overestimated value of $S$. The fluctuation of $S$ can be
explained by distortions of the droplet shape due to the dynamics of the
particle rotations. Another difference between the curved and the flat
interface is the time scale for the particle flip. However, one cannot see a
significant effect of $\chi$ on the time development of the order parameter.

To measure the spatial short range ordering of the particles, we define the
pair correlation function
\begin{equation}
 \label{eq:gd}
 g(r) = \frac{1}{2\pi\xi N} \left\langle\sum_{i, j \neq i} \int_{r-\frac{1}{2}}^{r+\frac{1}{2}} \delta(\tilde{r} - r_{ij})\, \text{d}\tilde{r}\right\rangle
\end{equation}
with discrete values for the distance $r$ between particle centres. The number
of particles in the system is given by $N$, and $\xi$ is a scaling factor
assuring that $g(r) \rightarrow1 $ for $r \rightarrow \infty$. Due to the
finite integration range, all pairs of particles with a distance $r_{ij} \in
[r - 1/2, r + 1/2]$ contribute to $g(r)$. In Fig.~\ref{fig:SandGr}(b), the
time dependence of $g(r)$ is shown for a spherical interface and $\chi =
0.305$. One notices that all maxima decrease in time. This means that the
local ordering decreases, particularly in the first $10^4$ time
steps. Afterwards, $g(r)$ changes only slightly. Fig.~\ref{fig:SandGr}(c)
compares $g(r)$ for $\chi = 0.153$ and $\chi = 0.305$ after $10^3$ time steps
and can be used to explain the differences of the snapshots in
Fig. \ref{fig:droplet_snapshots}(a) and (b): the order is higher for a smaller
concentration $\chi$. In the case of $\chi = 0.305$, the ellipsoids are
attracted by capillary forces leading to a clustering of particles. This
causes disorder, which manifests itself in smaller peak amplitudes of $g(r)$.

In conclusion, the particle coverage $\chi$ has a marginal influence on the
duration of the flipping towards the interface. However, the particles rotate
faster when the interface is spherical. For increasing $\chi$, the spatial
particle ordering is reduced since the average particle distance is
sufficiently small to allow for attractive capillary forces which in turn cause
particle clustering.

\subsubsection{Effects of particles on a droplet in shear flow}

In many industrial applications, the particle-covered droplets as discussed in the previous section are not stationary or in equilibrium, but are instead subjected to external stresses or forces. The properties of the individual droplets are then of interest as their behaviour dictates that of, for example, an emulsion formed of these droplets. In this example, we consider monodisperse neutrally wetting spherical particles. The particles are adsorped to the droplet interface which is initially spherical. As before, we employ the particle coverage fraction $\chi$: the fraction of the initial interfacial area taken away by the presence of these nanoparticles. In the simulations, variation of $\chi$ is achieved by varying the number of particles $N$ while keeping their radius $r_p$ constant. External shear is realised by using Lees-Edwards boundary conditions~\cite{bib:lees-edwards:1972}, adapted for use with lattice Boltzmann~\cite{bib:wagner-pagonabarraga:2002}. As this shear is applied to the system, the droplet will start to deform. To quantify this deformation, for small to moderate shear rates $\dot{\gamma}$, a dimensionless deformation parameter is used: $D = (L - B) / (L + B)$ where $L$ is the length and $B$ is the breadth of the droplet. This parameter will be zero for a sphere and tend to unity for a very strongly elongated droplet. As the droplet loses its spherical symmetry, it will also start to exhibit an angle of its long axis with respect to the shear flow: the inclination angle $\theta_d$.

\begin{figure}
 \subfloat[$\chi = 0.00$]{\shortstack{\includegraphics[width=0.33\linewidth]{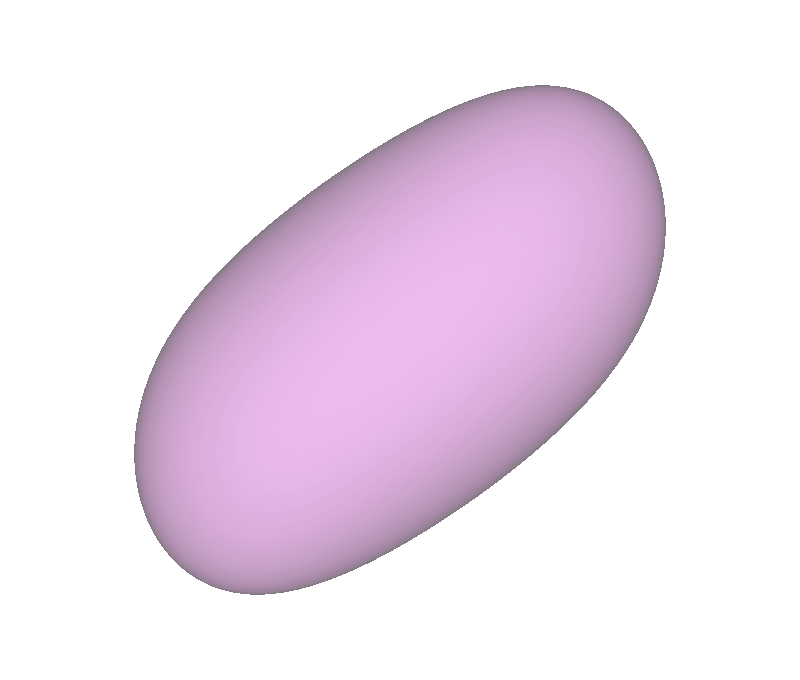} }}
 \subfloat[$\chi = 0.27$]{\hspace{-0.6cm}\shortstack{\includegraphics[width=0.33\linewidth]{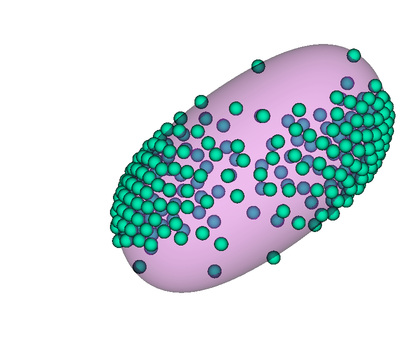} } \hspace{0.6cm}}
 \subfloat[$\chi = 0.55$]{\shortstack{\includegraphics[width=0.33\linewidth]{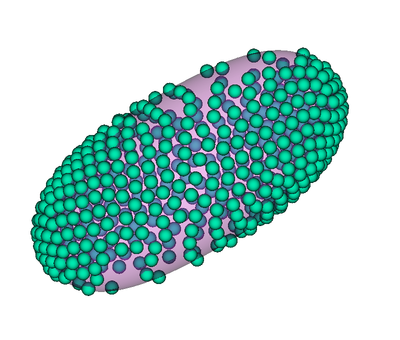} }} \\
 \subfloat[droplet deformation]{
  \includegraphics[width=0.475\linewidth]{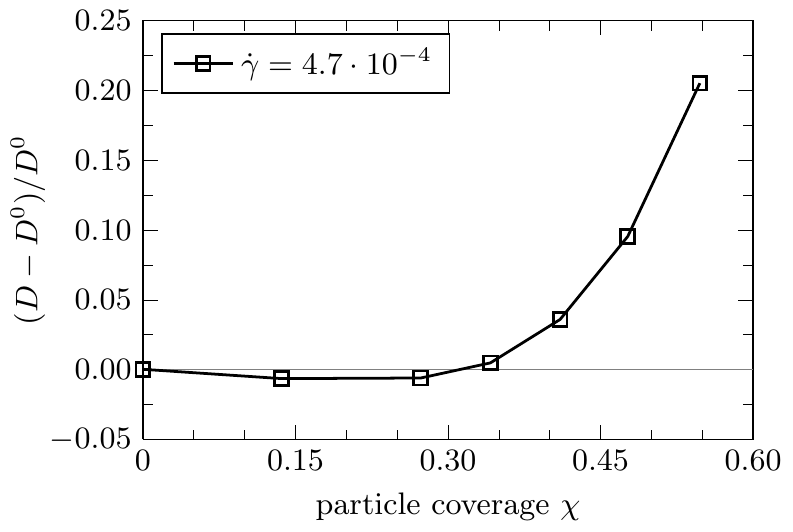}
 } \hfill
 \subfloat[droplet inclination]{
  \includegraphics[width=0.475\linewidth]{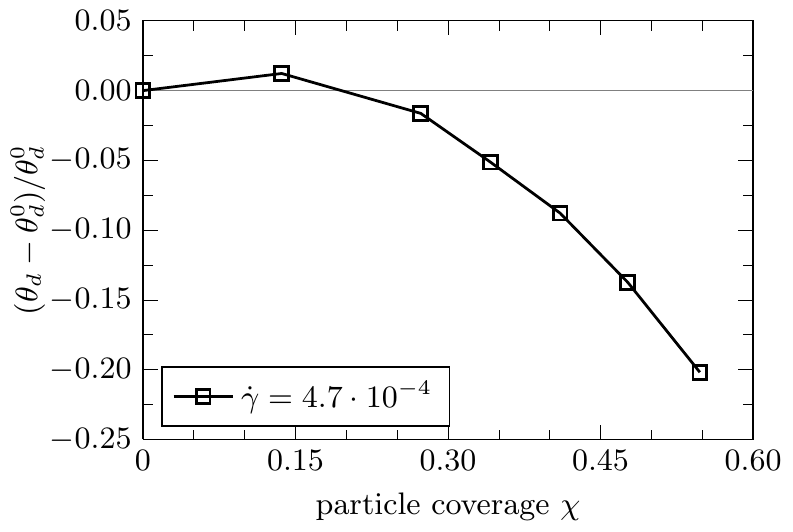}
}
 \caption{\label{fig:stefan-shear} Effect of adding monodisperse and
neutrally wetting spherical nanoparticles to the inferface of a droplet in
shear flow. Representative snapshots of deformed particle-covered droplets
are shown for $\dot{\gamma} = 4.7 \cdot 10^{-4}$ and (a) $\chi = 0$, (b)
$\chi = 0.27$, and (c) $\chi = 0.55$ (\cite{frijters2012effects},
reproduced with permission of the Royal Society of Chemistry). The relative
deformation of the droplet increases strongly for high coverage fraction,
as shown in (d). In turn, the relative inclination angle decreases,
indicating a better alignment with the shear flow, as shown in (e).}
\end{figure}

Increasing the shear rate a droplet is subjected to will generally increase its deformation, as well as reduce its inclination angle from an initial angle of 45 degrees: the droplet is elongated and aligns with the shear flow. However, this is only valid as long as inertia can be neglected. When inertial forces are comparable to or stronger than viscous forces, the inclination angle can first increase beyond 45 degrees, before its eventual reduction. We now discuss the effect of increasing the particle coverage fraction $\chi$ at constant shear rates $\dot{\gamma}$. Representative snapshots of the droplets for various $\chi$ and fixed $\dot{\gamma} = 0.47 \cdot 10^{-3}$ are presented in Fig.~\ref{fig:stefan-shear}(a), (b), and (c). The particles are not homogeneously distributed over the droplet surface when shear is applied, due to an interplay between local curvature and shear velocities: high curvatures and low velocities are energetically favoured.

Fig.~\ref{fig:stefan-shear}(d) shows the change in deformation of the
droplet as the particle coverage is increased. The deformation has been
rescaled as $D^* = (D - D^0) / D^0$, where $D^0 \equiv D(\chi = 0)$. Due
to a large decrease in free energy granted by the presence of particles at
the interface, they are irreversibly adsorbed. When shear is applied, the
particles are affected by this, and they start to move over the droplet
surface, but cannot be swept away. As can be seen from the figure, high
particle coverage fractions lead to a large increase in deformation of the
droplet. The inertia of the particles plays a critical role here. The
particles have to reverse direction to stay attached to the droplet, and
massive particles strongly resist this change in velocity. This causes the
interface to be dragged by the particles, increasing elongation from the
tips. Fig.~\ref{fig:stefan-shear}(e) demonstrates the effect of particle
coverage on the inclination angle of the droplet. Similar to the rescaling
performed on the deformation, the inclination angle has been rescaled as
$\theta_d^* = (\theta_d - \theta_d^0) / \theta_d^0$, where $\theta_d^0
\equiv \theta(\chi = 0)$. The inertial effects described above also cause
the inclination angle to strongly decrease for large $\chi$.

\section{Conclusions}
\label{sec:conclusions}

We have presented a non-exhaustive collection of simulation methods which are
able to catch the relevant physical ingredients to study the dynamics of
different kinds of fluid-fluid interfaces. The aim of this contribution is not
to provide all strengths and weaknesses of these methods, but to demonstrate
physical systems, where the selected methods are particularly suitable.
Obviously, numerous alternative simulation approaches exist, but it is beyond
the scope of this article to judge them by systematic quantitative benchmarks.

The immersed boundary method offers a flexible and simple approach to model
deformable particles immersed in suspending fluids, although it is known to be
somewhat problematic in the case of rigid boundaries, due to its explicit
algorithm. As it is a front-tracking method, the interface configuration is
directly known, which simplifies the force evaluation based on the interface
deformation. Another advantage is that the constitutive properties of the
interface can be controlled directly without tuning additional simulation
parameters, and there is no need to remesh the fluid domain. The IBM is especially suitable to simulate the dynamics of vesicles
and blood components at small and intermediate volume fractions.

The combination of the lattice Boltzmann method with the Shan-Chen
multi-component model for the simulation of immiscible fluids and a
lattice-based surfactant model allows to coarse-grain the molecular
interactions sufficiently well so as to reduce the computational effort
substantially. Thus, large system sizes can be reached allowing to study for
example macroscopic rheological properties and their dependence on microscopic
interactions. Furthermore, the model does not require any information of the
expected properties of a fluid mixture and the corresponding fluid-fluid
interfaces because all interactions are included on a mesoscopic level
providing a purely kinetic approach to the description of complex fluids. This
allows, for example, to study the formation of cubic mesophases such as the
gyroid phase without adding any information on the gyroid itself a priori.
However, a drawback of the model is that its phenomenological forces and
parameters are often difficult to be related to experimentally available
observables and time consuming parameter searches might be required to reveal
interesting physical behaviour.

When the multi-component lattice Boltzmann method is coupled to a molecular
dynamics algorithm for the description of suspended (colloidal) particles,
particle-laden interfaces can be studied at a level where not only
hydrodynamics, but also individual particles and their interactions are
resolved. At the same time, efficient scaling to allow larger domains to be
simulated is still provided due to the locality of the algorithm. While this
method has been proven to be particularly suitable to study particle stabilised
emulsions, one of its drawbacks is the diffuse interface between fluids in the
Shan-Chen LBM. The particles have to be substantially larger than these diffuse
interfaces in order to be able to stabilise them. Thus, even comparably simple
applications as the ones provided in this article require large lattices and
therefore access to supercomputing resources.

The models described above will gain complexity in the future. For example,
blood flow simulations will take into account molecular diffusion, as to account 
for messengers responsible for e.g. platelet activation. This requires elaborate
advection-diffusion-reaction models in the presence of complex and non-steady
boundaries. The investigation of particles on fluid-fluid interfaces will lead
to more general models as well: so far, in most studies, a series of potential
particle properties are ignored for the sake of simplicity (e.g., surface
charges and non-homogeneous surface chemistry, deformability, or more complex
shapes). The existing multi-component
lattice-Boltzmann models are also expected to see improvements in the future in
order to better account for specific interface properties and high density- or
viscosity ratios at moderate computational cost. All these extensions have to
be well-defined and therefore require tight collaborations with
experimentalists. The corresponding model developments will be rather
complicated and non-trivial, thus relying on contributions from theoreticians and
computer scientists.


\begin{thebibliography}{10}

\bibitem{seifert_configurations_1997}
U.~Seifert.
\newblock {Configurations of fluid membranes and vesicles}.
\newblock {\em Adv. Phys.}, 46:13--137, 1997.

\bibitem{pozrikidis2003modeling}
C.~Pozrikidis, editor.
\newblock {\em {Modeling and Simulation of Capsules and Biological Cells}}.
\newblock Chapman \& Hall/CRC Mathematical Biology and Medicine Series, 2003.

\bibitem{gompper2008soft}
G.~Gompper and M.~Schick.
\newblock {\em {Soft Matter: Lipid Bilayers and Red Blood Cells}}.
\newblock Wiley-VCH, 2008.

\bibitem{suresh2005connections}
S.~Suresh, J.~Spatz, J.~Mills, A.~Micoulet, M.~Dao, C.~Lim, M.~Beil, and
  M.~Seufferlein.
\newblock {Connections between single-cell biomechanics and human disease
  states: gastrointestinal cancer and malaria}.
\newblock {\em Acta Biomaterialia}, 1:15--30, 2005.

\bibitem{battaglia2012lipid}
L.~Battaglia and M.~Gallarate.
\newblock Lipid nanoparticles: state of the art, new preparation methods and
  challenges in drug delivery.
\newblock {\em Expert Opinion on Drug Delivery}, 9(5):497--508, 2012.

\bibitem{fedosov2011predicting}
D.~Fedosov, W.~Pan, B.~Caswell, G.~Gompper, and G.~Karniadakis.
\newblock Predicting human blood viscosity in silico.
\newblock {\em P. Natl. Acad. Sci.}, 108(29):11772, 2011.

\bibitem{hou2010deformability}
H.~Hou, A.~Bhagat, A.~Chong, P.~Mao, K.~Tan, J.~Han, and C.~Lim.
\newblock {Deformability based cell margination---A simple microfluidic design
  for malaria-infected erythrocyte separation}.
\newblock {\em Lab Chip}, 10(19):2605--2613, 2010.

\bibitem{silva2011nanoemulsions}
H.~Silva, M.~Cerqueira, and A.~Vicente.
\newblock Nanoemulsions for food applications: Development and
  characterization.
\newblock {\em Food and Bioprocess Technology}, pp 1--14, 2011.

\bibitem{puglia2012lipid}
C.~Puglia and F.~Bonina.
\newblock Lipid nanoparticles as novel delivery systems for cosmetics and
  dermal pharmaceuticals.
\newblock {\em Expert Opinion on Drug Delivery}, 9(4):429--441, 2012.

\bibitem{hirasaki2011recent}
G.~Hirasaki, C.~Miller, and M.~Puerto.
\newblock Recent advances in surfactant {EOR}.
\newblock {\em SPE Journal}, 16(4):889--907, 2011.

\bibitem{jafari2008re}
S.~Jafari, E.~Assadpoor, Y.~He, and B.~Bhandari.
\newblock Re-coalescence of emulsion droplets during high-energy
  emulsification.
\newblock {\em Food Hydrocolloids}, 22(7):1191--1202, 2008.

\bibitem{bib:jens-harvey-chin-venturoli-coveney:2005}
J.~Harting, M.~Harvey, J.~Chin, M.~Venturoli, and P.~V. Coveney.
\newblock Large-scale lattice {B}oltzmann simulations of complex fluids:
  advances through the advent of computational grids.
\newblock {\em {Phil. Trans. R. Soc. Lond. A}}, 363:1895--1915, 2005.

\bibitem{bib:jens-harvey-chin-coveney:2004}
J.~Harting, M.~Harvey, J.~Chin, and P.~V. Coveney.
\newblock Detection and tracking of defects in the gyroid mesophase.
\newblock {\em {Comp. Phys. Comm.}}, 165:97--109, 2004.

\bibitem{bib:gompper-schick:1994}
G.~Gompper and M.~Schick.
\newblock {\em Self-assembling amphiphilic systems}, volume~16.
\newblock Academic Press, 1994.

\bibitem{binks2002particles}
B.~Binks.
\newblock Particles as surfactants--similarities and differences.
\newblock {\em {Cur. Opin. Colloid Interface Sci.}}, 7(1-2):21--41, 2002.

\bibitem{tcholakova2008comparison}
S.~Tcholakova, N.~Denkov, and A.~Lips.
\newblock Comparison of solid particles, globular proteins and surfactants as
  emulsifiers.
\newblock {\em {Phys. Chem. Chem. Phys.}}, 10(12):1608--1627, 2008.

\bibitem{frijters2012effects}
S.~Frijters, F.~G{\"u}nther, and J.~Harting.
\newblock Effects of nanoparticles and surfactant on droplets in shear flow.
\newblock {\em {Soft Matter}}, 8(24):6542--6556, 2012.

\bibitem{bib:ramsden:1903}
W.~Ramsden.
\newblock Separation of solids in the surface-layers of solutions and
  `suspensions'.
\newblock {\em {Proc. R. Soc. London}}, 72:156, 1903.

\bibitem{pickering1907cxcvi}
S.~Pickering.
\newblock Emulsions.
\newblock {\em {J. Chem. Soc., Trans.}}, 91(0):2001--2021, 1907.

\bibitem{stratford2005colloidal}
K.~Stratford, R.~Adhikari, I.~Pagonabarraga, J.~Desplat, and M.~Cates.
\newblock Colloidal jamming at interfaces: A route to fluid-bicontinuous gels.
\newblock {\em Science}, 309(5744):2198--2201, 2005.

\bibitem{herzig2007bicontinuous}
E.~Herzig, K.~White, A.~Schofield, W.~Poon, and P.~Clegg.
\newblock Bicontinuous emulsions stabilized solely by colloidal particles.
\newblock {\em {Nature Materials}}, 6(12):966--971, 2007.

\bibitem{kim2010bijels}
E.~Kim, K.~Stratford, and M.~Cates.
\newblock Bijels containing magnetic particles: A simulation study.
\newblock {\em {Langmuir}}, 26(11):7928--7936, 2010.

\bibitem{binks2001particles}
B.~Binks and P.~Fletcher.
\newblock Particles adsorbed at the oil-water interface: A theoretical
  comparison between spheres of uniform wettability and {J}anus particles.
\newblock {\em Langmuir}, 17(16):4708--4710, 2001.

\bibitem{sagis2011dynamic}
L.~Sagis.
\newblock {Dynamic properties of interfaces in soft matter: Experiments and
  theory}.
\newblock {\em {Rev. Mod. Phys.}}, 83(4):1367, 2011.

\bibitem{barths-biesel_time-dependent_1981}
D.~Barth{\`e}s-Biesel and J.~Rallison.
\newblock {The Time-Dependent Deformation of a Capsule Freely Suspended in a
  Linear Shear Flow}.
\newblock {\em {J. Fluid Mech.}}, 113:251--267, 1981.

\bibitem{wetzel2001droplet}
E.~Wetzel.
\newblock Droplet deformation in dispersions with unequal viscosities and zero
  interfacial tension.
\newblock {\em {J. Fluid Mech.}}, 426:199--228, 2001.

\bibitem{larson1999structure}
R.~Larson.
\newblock {\em {The Structure and Rheology of Complex Fluids}}.
\newblock Oxford University Press, 1999.

\bibitem{benzi2010herschel}
R.~Benzi, M.~Bernaschi, M.~Sbragaglia, and S.~Succi.
\newblock {Herschel-Bulkley rheology from lattice kinetic theory of soft glassy
  materials}.
\newblock {\em {Europhys. Lett.}}, 91:14003, 2010.

\bibitem{pal2011rheology}
R.~Pal.
\newblock {Rheology of simple and multiple emulsions}.
\newblock {\em Curr. Opin. Colloid In.}, 16:41--60, 2011.

\bibitem{Succi2001}
S.~Succi.
\newblock {\em The Lattice Boltzmann Equation}.
\newblock Oxford University Press, Oxford, UK, 2001.

\bibitem{Sukop2006}
M.~Sukop and D.~Thorne.
\newblock {\em Lattice Boltzmann Modeling: An Introduction for Geoscientists
  and Engineers}.
\newblock Springer, Berlin, Germany, 2006.

\bibitem{Aidun2010}
C.~Aidun and J.~Clausen.
\newblock Lattice-{B}oltzmann method for complex flows.
\newblock {\em Annu. Rev. Fluid Mech}, 42:439, 2010.

\bibitem{qian_lattice_1992}
Y.-H. Qian, D.~{d'Humi{\`e}res}, and P.~Lallemand.
\newblock {Lattice BGK Models for Navier-Stokes Equation}.
\newblock {\em Europhys. Lett.}, 17:479--484, 1992.

\bibitem{Shan1993}
X.~Shan and H.~Chen.
\newblock Lattice {B}oltzmann model for simulating flows with multiple phases
  and components.
\newblock {\em Phys. Rev. E}, 47:1815, 1993.

\bibitem{bib:shan-chen:1994}
X.~Shan and H.~Chen.
\newblock Simulation of nonideal gases and liquid-gas phase transitions by the
  lattice {B}oltzmann equation.
\newblock {\em {Phys. Rev. E}}, 49:2941, 1994.

\bibitem{bib:orlandini-swift-yeomans:1995}
E.~Orlandini, M.~R. Swift, and J.~M. Yeomans.
\newblock A lattice {B}oltzmann model of binary-fluid mixtures.
\newblock {\em {Europhys. Lett.}}, 32:463, 1995.

\bibitem{bib:swift-orlandini-osborn-yeomans:1996}
M.~R. Swift, E.~Orlandini, W.~R. Osborn, and J.~M. Yeomans.
\newblock Lattice-{B}oltzmann simulations of liquid-gas and binary fluid
  systems.
\newblock {\em {Phys. Rev. E}}, 54:5041, 1996.

\bibitem{bib:dupin-halliday-care:2003}
M.~Dupin, I.~Halliday, and C.~Care.
\newblock Multi-component lattice {B}oltzmann equation for mesoscale blood
  flow.
\newblock {\em J. Phys. A: Math. Gen.}, 36:8517, 2003.

\bibitem{bib:lishchuk-care-halliday:2003}
S.~Lishchuk, C.~Care, and I.~Halliday.
\newblock Lattice {B}oltzmann algorithm for surface tension with greatly
  reduced microcurrents.
\newblock {\em {Phys. Rev. E}}, 67:036701, 2003.

\bibitem{Peskin2002}
C.~Peskin.
\newblock The immersed boundary method.
\newblock {\em Acta Numerica}, 11:479, 2002.

\bibitem{tryggvason_front-tracking_2001}
G.~Tryggvason, B.~Bunner, A.~Esmaeeli, D.~Juric, N.~Al-Rawahi, W.~Tauber,
  J.~Han, S.~Nas, and Y.-J. Jan.
\newblock {A front-tracking method for the computations of multiphase flow}.
\newblock {\em {J. Comput. Phys.}}, 169(2):708--759, 2001.

\bibitem{kruger2011efficient}
T.~Kr{\"u}ger, F.~Varnik, and D.~Raabe.
\newblock {Efficient and accurate simulations of deformable particles immersed
  in a fluid using a combined immersed boundary lattice Boltzmann finite
  element method.}
\newblock {\em Comput. Math. Appl.}, 61:3485--3505, 2011.

\bibitem{bib:chen-lookman:1995}
S.~Chen and T.~Lookman.
\newblock Growth kinetics in multicomponent fluids.
\newblock {\em {J. Stat. Phys.}}, 81:223, 1995.

\bibitem{bib:chen-boghosian-coveney-nekovee:2000}
H.~Chen, B.~Boghosian, P.~Coveney, and M.~Nekovee.
\newblock A ternary lattice {B}oltzmann model for amphiphilic fluids.
\newblock {\em {Proc. R. Soc. Lond. A}}, 456:2043, 2000.

\bibitem{bib:nekovee-coveney-chen-boghosian:2000}
M.~Nekovee, P.~Coveney, H.~Chen, and B.~Boghosian.
\newblock Lattice-{B}oltzmann model for interacting amphiphilic fluids.
\newblock {\em {Phys. Rev. E}}, 62:8282, 2000.

\bibitem{bib:furtado-skartlien:2010}
K.~Furtado and R.~Skartlien.
\newblock Derivation and thermodynamics of a lattice {B}oltzmann model with
  soluble amphiphilic surfactant.
\newblock {\em {Phys. Rev. E}}, 81:066704, 2010.

\bibitem{bib:lamura-gonnella-yeomans:1999}
A.~Lamura, G.~Gonnella, and J.~M. Yeomans.
\newblock A lattice {B}oltzmann model of ternary fluid mixtures.
\newblock {\em {Europhys. Lett.}}, 45:314, 1999.

\bibitem{bib:benzi-chibbaro-succi:2009}
R.~Benzi, S.~Chibbaro, and S.~Succi.
\newblock Mesoscopic lattice {B}oltzmann modeling of flowing soft systems.
\newblock {\em {Phys. Rev. Lett.}}, 102:026002, 2009.

\bibitem{bib:benzi-sbragaglia-succi-bernaschi-chibbaro:2009}
R.~Benzi, M.~Sbragaglia, S.~Succi, M.~Bernaschi, and S.~Chibbaro.
\newblock Mesoscopic lattice {B}oltzmann modeling of soft-glassy systems:
  Theory and simulations.
\newblock {\em {J. Chem. Phys.}}, 131:104903, 2009.

\bibitem{jansen2011bijels}
F.~Jansen and J.~Harting.
\newblock {From bijels to Pickering emulsions: A lattice Boltzmann study}.
\newblock {\em {Phys. Rev. E}}, 83(4):046707, 2011.

\bibitem{Guenther2012a}
F.~G{\"u}nther, F.~Janoschek, S.~Frijters, and J.~Harting.
\newblock Lattice {B}oltzmann simulations of anisotropic particles at liquid
  interfaces.
\newblock {\em Comput. Fluids}, 2012.

\bibitem{bib:joshi-sun:2010}
A.~Joshi and Y.~Sun.
\newblock Wetting dynamics and particle deposition for an evaporating colloidal
  drop: A lattice {B}oltzmann study.
\newblock {\em {Phys. Rev. E}}, 82:041401, 2010.

\bibitem{bib:joshi-sun:2009}
A.~Joshi and Y.~Sun.
\newblock Multiphase lattice {B}oltzmann method for particle suspensions.
\newblock {\em {Phys. Rev. E}}, 79:066703, 2009.

\bibitem{Ladd2001}
A.~Ladd and R.~Verberg.
\newblock Lattice-{B}oltzmann simulations of particle-fluid suspensions.
\newblock {\em J. Stat. Phys}, 104:1191, 2001.

\bibitem{Nguyen2002}
N.-Q. Nguyen and A.~Ladd.
\newblock Lubrication correction for lattice-{B}oltzmann simulation of particle
  suspensions.
\newblock {\em Phys. Rev. E}, 66:046708, 2002.

\bibitem{Hertz1881}
H.~Hertz.
\newblock {\"U}ber die {B}er{\"u}hrung fester elastischer {K}{\"o}rper.
\newblock {\em Journal f{\"u}r reine und angewandte Mathematik}, 92:156, 1881.

\bibitem{Berne1972}
B.~Berne and P.~Pechukas.
\newblock Gaussian model potentials for molecular interactions.
\newblock {\em J. Chem. Phys.}, 56:4213, 1972.

\bibitem{Keller1982}
S.~Keller and R.~Skalak.
\newblock Motion of a tank-treading ellipsoidal particle in a shear flow.
\newblock {\em J. Fluid Mech}, 120:27, 1982.

\bibitem{Mader2006}
M.~Mader, V.~Vitkova, M.~Abkarian, A.~Viallat, and T.~Podgorski.
\newblock Dynamics of viscous vesicles in shear flow.
\newblock {\em Eur. Phys. J. E.}, 19:389397, 2006.

\bibitem{Kantsler2006}
V.~Kantsler and V.~Steinberg.
\newblock Transition to tumbling and two regimes of tumbling motion of a
  vesicle in shear flow.
\newblock {\em Phys. Rev. Lett}, 96:036001, 2006.

\bibitem{Kaoui2011a}
B.~Kaoui, J.~Harting, and C.~Misbah.
\newblock Two-dimensional vesicle dynamics under shear flow: Effect of
  confinement.
\newblock {\em Phys. Rev. E}, 83:066319, 2011.

\bibitem{Kaoui2012}
B.~Kaoui, T.~Kr{\"u}ger, and J.~Harting.
\newblock How does confinement affect the dynamics of viscous vesicles and red
  blood cells?
\newblock {\em Soft Matter}, (doi:10.1039/C2SM26289D), 2012.

\bibitem{Kaoui2008}
B.~Kaoui, G.~Ristow, I.~Cantat, C.~Misbah, and W.~Zimmermann.
\newblock Lateral migration of a two-dimensional vesicle in unbounded
  {P}oiseuille flow.
\newblock {\em Phys. Rev. E}, 77:021903, 2008.

\bibitem{geislinger2012separation}
T.~Geislinger, B.~Eggart, S.~Braunm{\"u}ller, L.~Schmid, and T.~Franke.
\newblock Separation of blood cells using hydrodynamic lift.
\newblock {\em Appl. Phys. Lett.}, 100(18):183701, 2012.

\bibitem{kruger2011particlestress}
T.~Kr{\"u}ger, F.~Varnik, and D.~Raabe.
\newblock {Particle stress in suspensions of soft objects}.
\newblock {\em {Phil. Trans. R. Soc. Lond. A}}, 369:2414--2421, 2011.

\bibitem{saksena-coveney-2008}
R.~S. Saksena and P.~V. Coveney.
\newblock Self-assembly of ternary cubic, hexagonal, and lamellar mesophases
  using the lattice-{B}oltzmann kinetic method.
\newblock {\em J. Phys. Chem. B}, 112:2950, 2008.

\bibitem{bib:gonzalez-nekovee-coveney}
N.~Gonz\'{a}lez-Segredo, M.~Nekovee, and P.~V. Coveney.
\newblock Three-dimensional lattice-{B}oltzmann simulations of critical
  spinodal decomposition in binary immiscible fluids.
\newblock {\em Phys. Rev. E}, 67:046304, 2003.

\bibitem{bib:jens-giupponi-coveney:2006}
G.~Giupponi, J.~Harting, and P.~V. Coveney.
\newblock Emergence of rheological properties in lattice {B}oltzmann
  simulations of gyroid mesophases.
\newblock {\em Europhys. Lett.}, 73:533, 2006.

\bibitem{bib:emerton-coveney-boghosian}
A.~N. Emerton, P.~V. Coveney, and B.~M. Boghosian.
\newblock Lattice-gas simulations of domain growth, saturation and
  self-assembly in immiscible fluids and microemulsions.
\newblock {\em Phys. Rev. E}, 56(1):1286, 1997.

\bibitem{bib:gonzalez-coveney-2}
N.~Gonz\'{a}lez-Segredo and P.~V. Coveney.
\newblock Coarsening dynamics of ternary amphiphilic fluids and the
  self-assembly of the gyroid and sponge mesophases: lattice-{B}oltzmann
  simulations.
\newblock {\em Phys. Rev. E}, 69:061501, 2004.

\bibitem{bib:jens-giupponi-coveney:2007}
J.~Harting, G.~Giupponi, and P.~V. Coveney.
\newblock Structural transitions and arrest of domain growth in sheared binary
  immiscible fluids and microemulsions.
\newblock {\em Phys. Rev. E}, 75:041504, 2007.

\bibitem{bib:gonzalez-coveney}
N.~Gonz\'{a}lez-Segredo and P.~V. Coveney.
\newblock Self-assembly of the gyroid cubic mesophase: lattice-{B}oltzmann
  simulations.
\newblock {\em Europhys. Lett.}, 65(6):795--801, 2004.

\bibitem{bib:gandy-klinowski}
P.~J.~F. Gandy and J.~Klinowski.
\newblock Exact computation of the triply periodic g ('gyroid') minimal
  surface.
\newblock {\em Chem. Phys. Lett.}, 321(5):363, 2000.

\bibitem{bib:grosse-brauckmann}
K.~Gro{\ss}e-Brauckmann.
\newblock Gyroids of constant mean curvature.
\newblock {\em Exp. Math.}, 6(1):33, 1997.

\bibitem{bib:chan-hoffman-etal}
V.~Z.~H. Chan, J.~Hoffman, V.~Y. Lee, H.~Iatrou, A.~Avgeropoulos,
  N.~Hadjichristidis, R.~D. Miller, and E.~L. Thomas.
\newblock Ordered bicontinuous nanoporous and nanorelief ceramic films from
  self assembling polymer precursors.
\newblock {\em Science}, 286(5445):1716, 1999.

\bibitem{saksena-coveney-2009}
R.~S. Saksena and P.~V. Coveney.
\newblock Shear rheology of amphiphilic cubic liquid crystals from large-scale
  kinetic lattice-{B}oltzmann simulations.
\newblock {\em Soft Matter}, 5:4446, 2009.

\bibitem{Kaoui2011c}
A.~Ghosh, J.~Harting, M.~van Hecke, A.~Siemens, B.~Kaoui, V.~Koning,
  K.~Langner, I.~Niessen, J.~P. Rojas, S.~Stoyanov, and M.~Dijkstra.
\newblock Structuring with anisotropic colloids.
\newblock In {\em Proceedings of the Workshop Physics with Industry (Leiden,
  The Netherlands, October 17-21, 2011)}. Stichting FOM, 2011.

\bibitem{Zeng2012a}
C.~Zeng, F.~Brau, B.~Davidovitch, and A.~D. Dinsmore.
\newblock Capillary interactions among spherical particles at curved liquid
  interfaces.
\newblock {\em Soft Matter}, 8:8582--8594, 2012.

\bibitem{bib:lees-edwards:1972}
A.~Lees and S.~Edwards.
\newblock The computer study of transport processes under extreme conditions.
\newblock {\em {J. Phys. C.}}, 5:1921, 1972.

\bibitem{bib:wagner-pagonabarraga:2002}
A.~Wagner and I.~Pagonabarraga.
\newblock Lees-{E}dwards boundary conditions for lattice {B}oltzmann.
\newblock {\em {J. Stat. Phys.}}, 107:521, 2002.

\end{thebibliography}

\end{document}